\documentclass[conference]{IEEEtran}
\IEEEoverridecommandlockouts

\makeatletter
\newcommand{\linebreakand}{%
  \end{@IEEEauthorhalign}
  \hfill\mbox{}\par
  \mbox{}\hfill\begin{@IEEEauthorhalign}
}
\makeatother

\usepackage{url}
\usepackage{amsmath,amsfonts}
\usepackage{algorithmic}
\usepackage{graphicx}
\usepackage{textcomp}
\usepackage{xcolor}
\usepackage[ruled,linesnumbered,vlined]{algorithm2e}
\usepackage{pifont}
\usepackage{caption}
\usepackage{subcaption}
\usepackage{xspace}
\usepackage{multirow}
\usepackage{cleveref}
\usepackage{siunitx}
\usepackage{enumitem}
\usepackage{booktabs}
\usepackage{makecell}
\usepackage{graphicx}
\setlist[itemize]{topsep=2pt}

\def\BibTeX{{\rm B\kern-.05em{\sc i\kern-.025em b}\kern-.08em
    T\kern-.1667em\lower.7ex\hbox{E}\kern-.125emX}}
    
\begin{document}

\pdfpagewidth=8.5in
\pdfpageheight=11in

% submission number
\newcommand{\iscasubmissionnumber}{1360}
% project macro
\newcommand*{\project}{\texttt{SLOTH}\xspace}

\pagenumbering{arabic}

\title{\project: Lightweight Detection and Localization of On-Chip Fail-Slow Failures for DNN Accelerators}

% \author{
%   Junchi Wu \\
%   Peking University \\
%   Beijing, China \\
%   2501111908@stu.pku.edu.cn
%   \and
%   Xinfei Wan \\
%   Peking University \\
%   Beijing, China \\
%   2300013096@stu.pku.edu.cn
%   \and
%   Zhuoran Li \\
%   Peking University \\
%   Beijing, China \\
%   2200012710@stu.pku.edu.cn
%   \and
%   Yuyang Jin \\
%   Tsinghua University \\
%   Beijing, China \\
%   jyy17@mails.tsinghua.edu.cn
%   \and
%   Guangyu Sun \\
%   Peking University \\
%   Beijing, China \\
%   gsun@pku.edu.cn
%   \and
%   Yun Liang \\
%   Peking University \\
%   Beijing, China \\
%   ericlyun@pku.edu.cn
%   \and
%   Diyu Zhou \\
%   Peking University \\
%   Beijing, China \\
%   diyu.zhou@pku.edu.cn
%   \and
%   Youwei Zhuo \\
%   Peking University \\
%   Beijing, China \\
%   youwei@pku.edu.cn
% }

\author{
    % --- 第一排 (3人) ---
    \IEEEauthorblockN{Junchi Wu}
    \IEEEauthorblockA{Peking University\\
    Beijing, China\\
    2501111908@stu.pku.edu.cn}
    \and
    \IEEEauthorblockN{Xinfei Wan}
    \IEEEauthorblockA{Peking University\\
    Beijing, China\\
    2300013096@stu.pku.edu.cn}
    \and
    \IEEEauthorblockN{Zhuoran Li}
    \IEEEauthorblockA{Peking University\\
    Beijing, China\\
    2200012710@stu.pku.edu.cn}
    \linebreakand % <--- 强制换行
    % --- 第二排 (3人) ---
    \IEEEauthorblockN{Yuyang Jin}
    \IEEEauthorblockA{Tsinghua University\\
    Beijing, China\\
    jyy17@mails.tsinghua.edu.cn}
    \and
    \IEEEauthorblockN{Guangyu Sun}
    \IEEEauthorblockA{Peking University\\
    Beijing, China\\
    gsun@pku.edu.cn}
    \and
    \IEEEauthorblockN{Yun Liang}
    \IEEEauthorblockA{Peking University\\
    Beijing, China\\
    ericlyun@pku.edu.cn}
    \linebreakand % <--- 强制换行
    % --- 第三排 (2人) ---
    \IEEEauthorblockN{Diyu Zhou}
    \IEEEauthorblockA{Peking University\\
    Beijing, China\\
    diyu.zhou@pku.edu.cn}
    \and
    \IEEEauthorblockN{Youwei Zhuo}
    \IEEEauthorblockA{Peking University\\
    Beijing, China\\
    youwei@pku.edu.cn}
}
    
\maketitle
\thispagestyle{plain}
\pagestyle{plain}

% sections
\begin{abstract}

Spatial DNN accelerators are essential for high-performance inference, but their performance is undermined by widespread fail-slow failures. Detecting such failures on-chip is challenging, as prior methods from distributed systems are unsuitable due to strict memory limits and their inability to track failures across the hardware topology.
We present \project, a lightweight, hardware-aware framework for practical on-chip fail-slow detection in DNN accelerators.
\project combines workload-aware instrumentation for operator-level monitoring with minimal overhead, on-the-fly trace compression to operate within kilobytes of memory, and a novel topology-aware ranking algorithm to pinpoint a failure's root cause.
We evaluate \project on a wide range of representative DNN workloads. The results demonstrate that \project reduces the storage overhead by an average of 115.9$\times$, while achieving an average fail-slow detection accuracy from 69.68\% to 86.69\%.

\end{abstract}

\section{Introduction}

A \textbf{fail-slow failure} occurs when a hardware component remains functionally correct but operates at significantly reduced speed~\cite{gunawi2018fail}. Unlike fail-stop faults that halt execution outright~\cite{pillai2017application,alagappan2016correlated}, fail-slow faults let the system continue producing correct results, so degradation accumulates undetected. These faults are pervasive and costly in practice: in large GPU clusters, over 40\% of training jobs experience straggling, wasting more than 10\% of allocated GPU-hours, causing task completion time delayed 35\% on average~\cite{lin2025understanding,wu2024falconpinpointingmitigatingstragglers}; even a small fraction of persistently degraded nodes can reduce large-job completion rates by 30\%~\cite{tyrrell2024revisiting}.

\begin{figure}[t]
\centering
\includegraphics[width=\linewidth]{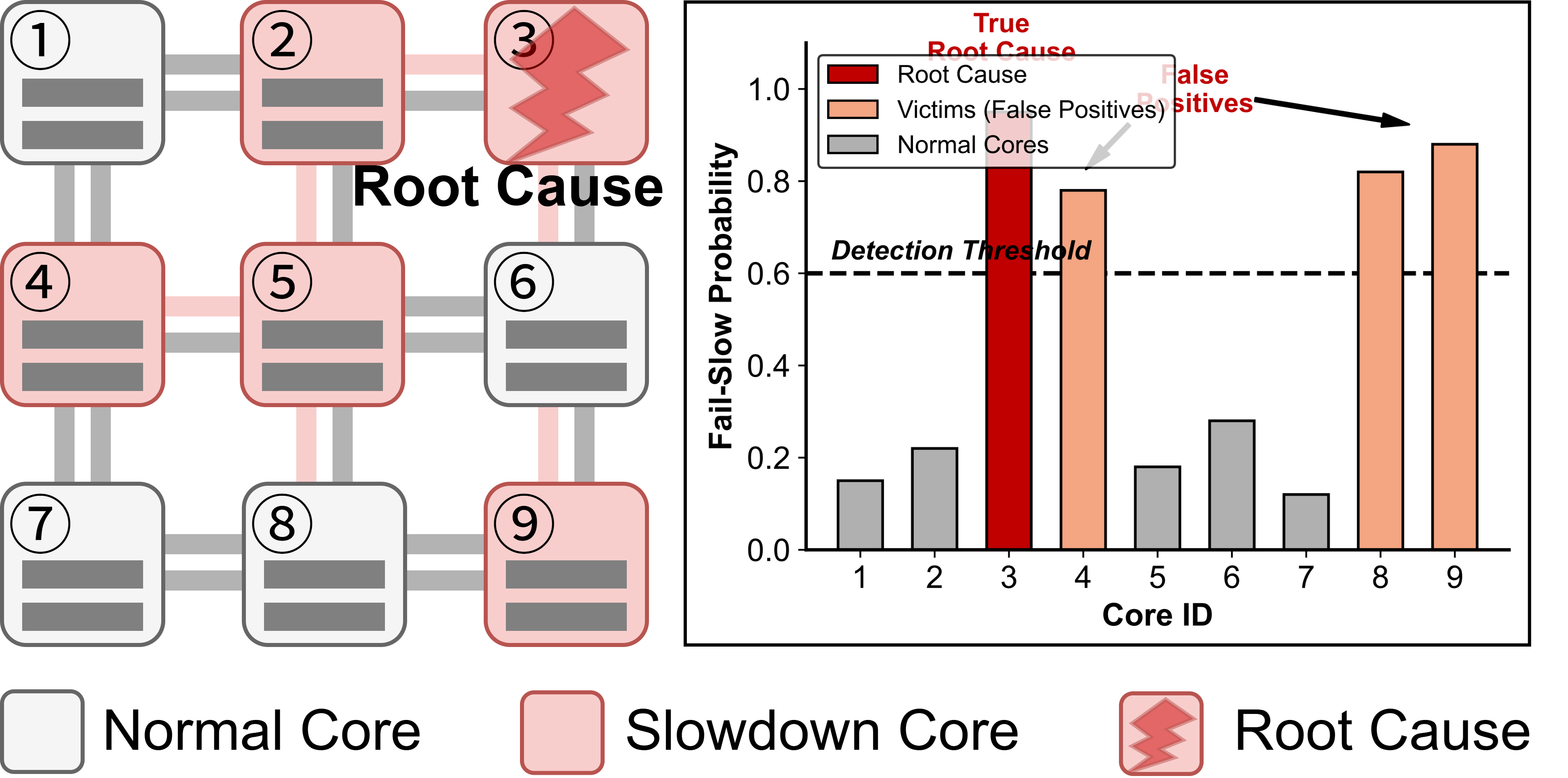}
\captionsetup{skip=0pt}
\caption{Per-core granularity detection: trouble distinguishing the true root cause from its propagated victim cores, leading to widespread false positives.
}
\label{fig:slowdown}
\vspace{-15pt}
\end{figure}

Yet current detection techniques, developed for distributed and cloud
systems~\cite{10.5555/3768039.3768105,10.5555/3585938.3585942,10.5555/3767955.3767975,groot,sleuth,panda2019iaso},
operate at \textbf{node or server granularity}:
they can flag a slow machine, but not which internal component is
responsible.
As shown in Figure~\ref{fig:slowdown}, traditional per-core granularity method fails to detect the root cause accurately.
Within a DNN accelerator, where hundreds to thousands of
cores are interconnected by a
Network-on-Chip~(NoC)~\cite{10.1145/3581784.3627042,kundu2025comparisoncerebraswaferscaleintegration,10485099,7056026},
individual component degradation is \textbf{statistically inevitable}, and this coarse granularity becomes insufficient: treating an entire chip as a
single monolithic node and replacing it upon any performance
drop is no longer economically viable.
As a result, fine-grained detection that localizes faults to \textbf{individual on-chip cores and links} remains an open problem.

The need for such detection is especially acute in spatial DNN accelerators. 
These architectures distribute operators across an array of cores connected by an on-chip network; each operator fires as soon as its inputs arrive, creating tight coupling between cores. 
This tightly coupled dataflow relies heavily on pipeline buffers and dataflow synchronization, which could create an \textbf{observability gap}: when a core experiences a minor fail-slow, its initial latency is often absorbed by downstream hardware queues or masked by synchronization waits. 
Consequently, these micro-level fail-slows are \textbf{hidden from external interfaces.} Compounding the problem, fail-slow faults are often transient, appearing under specific workload or thermal conditions~\cite{gunawi2018fail}; 
by the time the delay finally saturates the buffers and cascades across the chip to become visible to coarse-grained external monitors, the root cause may have already vanished. 
Overcoming this gap with fine-grained, intra-chip localization \textbf{unlocks new mitigation strategies:} rerouting traffic or remapping workloads around the degraded component rather than taking the entire chip offline.

\begin{figure}[tbp]
  \centering
  \begin{minipage}[b]{0.44\linewidth}
    \centering
    \includegraphics[width=\linewidth]{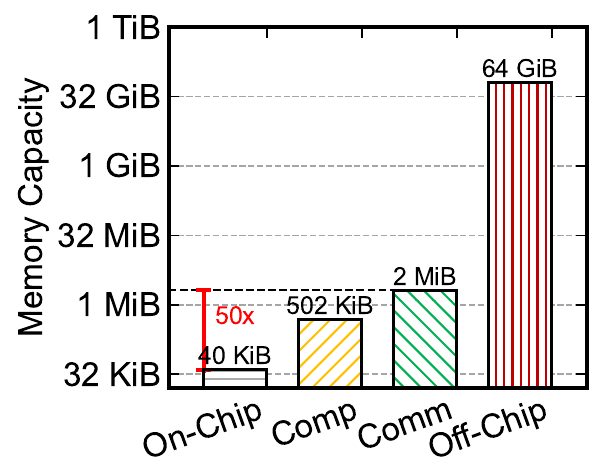}
    \label{fig:memory}
  \end{minipage}
  \hfill
  \begin{minipage}[b]{0.54\linewidth}
    \centering
    \includegraphics[width=\linewidth]{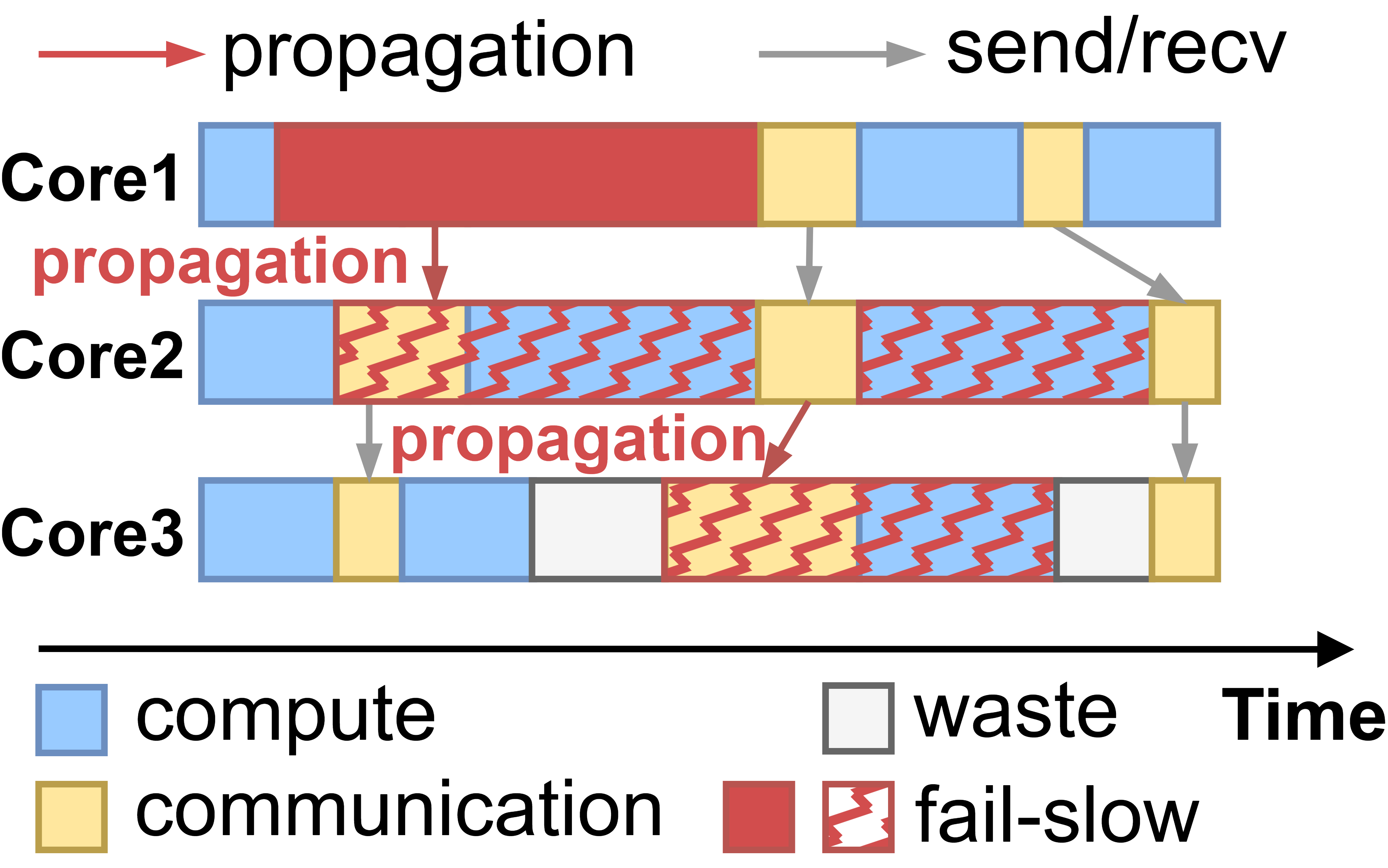}
    \label{fig:propagate}
  \end{minipage}
  \captionsetup{skip=-2pt}
  \caption{Challenges for detecting fail-slow failures: (a) On-chip, off-chip memory capacity versus computation and communication tracing log size for tracing DarkNet-19 for 1 second on a 4$\times$4 core processor. (b) Slowdown propagates across cores along communication dependencies. }
  \label{fig:memory&propagate}
  \vspace{-20pt}
\end{figure}

However, detecting fail-slow faults at this on-chip granularity
faces two obstacles.
First, \textbf{on-chip memory is severely constrained}. Detectors
designed for cloud and storage systems, such as
Sieve~\cite{10.5555/3768039.3768105} and
Perseus~\cite{10.5555/3585938.3585942}, rely on system-level logs or
per-request latency traces requiring megabytes to gigabytes of storage.
As shown in Figure~\ref{fig:memory&propagate}a, even a minimal
1-second communication trace of DarkNet-19 requires
\SI{2}{\mebi\byte}, far exceeding the typical on-chip SRAM capacity
of no more than \SI{64}{\kibi\byte} per core.
Second, \textbf{these methods do not model the hardware topology}.
Graph-based root-cause analyzers such as Groot~\cite{groot} and
Sleuth~\cite{sleuth} construct service dependency graphs from
distributed traces, while ADR~\cite{10.5555/3767955.3767975} relies on
host-level or switch-level telemetry; neither models the physical NoC
interconnect. As shown in Figure~\ref{fig:memory&propagate}b, when a
fail-slow in Core~1 delays its computation, the slowdown propagates
along the communication path: downstream Cores~2 and~3 are forced into
wait states and exhibit fail-slow-like symptoms themselves. Without
modeling the NoC topology, these methods cannot distinguish
backpressure victims from the actual culprit.

An effective on-chip detection mechanism must therefore be
{lightweight}, operating within kilobytes of memory rather
than gigabytes, and {topology-aware}, reasoning over the
physical NoC structure to separate root causes from propagated
effects.
In this paper, we present
\textbf{\project}\footnote{The source code of \project along with
failure datasets and traces are available at
\url{https://anonymous.4open.science/r/sloth-91D7}}(\textbf{S}lowdown
\textbf{L}ocalization via \textbf{O}n-chip \textbf{T}racing and
\textbf{H}ierarchical graph analysis), a framework that combines
lightweight on-chip tracing with topology-aware root cause analysis to
close this gap. To the best of our knowledge, \project is the first framework that addresses fail-slow detection at the on-chip DNN accelerator level. Unlike prior detection methods, \project distinguishes itself by its hardware topology-awareness, which performs reasoning directly on the extended topology graph. In summary, our contributions are:

\begin{itemize}
\item \textbf{SL-Compiler}: A compile-time instrumentation mechanism
that analyzes the workload's computation graph to place timing probes
at the most informative operator boundaries, enabling per-operator
timing collection at less than 10\% runtime overhead.
\item \textbf{SL-Recorder}: A two-stage online compression
structure that reduces trace storage by over 100$\times$ within each
core's kilobyte-scale SRAM. The first stage groups operators by
workload pattern; the second maintains per-bucket latency statistics
to retain the rare deviations that signal fail-slow onset
while discarding steady-state redundancy.
\item \textbf{SL-Tracer}: A diagnosis pipeline built on a
multi-level communication graph that fuses software dataflow
dependencies with the physical NoC topology. On this graph, FailRank,
an iterative scoring algorithm, traces backpressure propagation to
separate true root causes from cascading symptoms, localizing faults
to individual cores or links.
\end{itemize}

We evaluate \project on a cycle-level
simulator validated against the Dataflow Abstract Machine (DAM)
framework~\cite{10.1109/ISCA59077.2024.00046}.
To model realistic failure scenarios, we construct two
fault datasets grounded in real-world failure mechanisms: a
spatial-pattern dataset derived from production wafer map defect
distributions~\cite{wu2014wafer}, and a workload-thermal dataset that
models utilization-driven hotspots based on per-core activity
profiling~\cite{Srinivasan2003RAMPA,gunawi2018fail}. Against five
baseline methods,
\project achieves an average fail-slow detection accuracy from 69.68\% to 86.69\%, reduces trace storage by 115.9$\times$.

\section{Background}

\subsection{Spatial DNN Accelerators}

\begin{figure}[tbp]
  \centering
  \includegraphics[width=\linewidth]{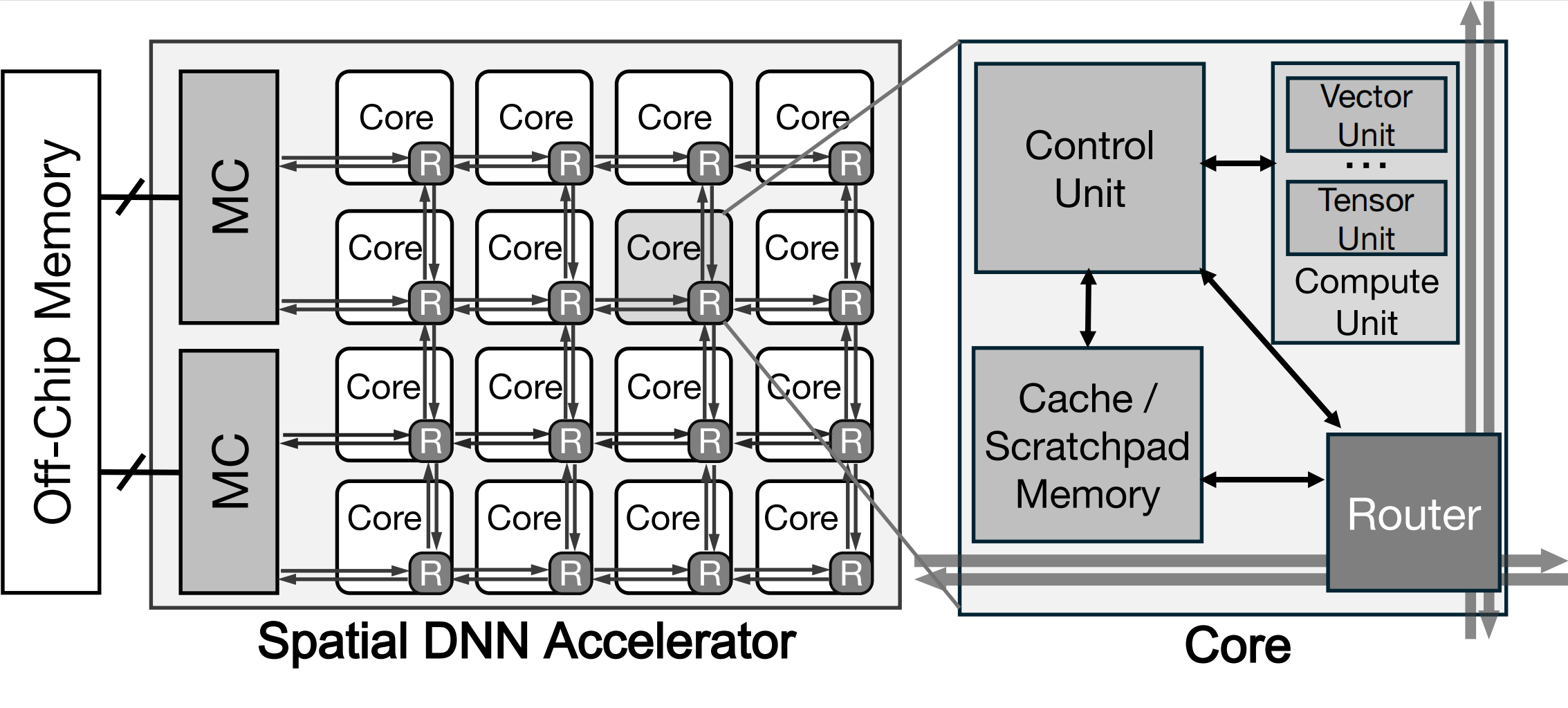}
  \caption{Overview of a spatial DNN accelerator with 2D-mesh network and its intra-tile organization.}
  \label{fig:manycore}
  \vspace{-12pt}
\end{figure}

Spatial DNN accelerators integrate an array of cores into a scalable on-chip network to exploit parallelism in deep learning inference~\cite{chen2016eyeriss,lie2023cerebras,cai2024gemini,prabhakar2024sambanova,genc2024stellar}.
This class of domain-specific accelerator is the hardware target of this work.
Figure \ref{fig:manycore} illustrates the fundamental components of such an architecture. Inter-core communication is enabled by the Network-on-Chip (NoC), which can adopt various topologies. Common structures include 2D-mesh, torus, and ring networks \cite{zheng2021adapt,feng2024ring}, while more complex topologies such as hypercube and dragonfly have also been proposed \cite{feng2023scalable}. These topologies can support different communication patterns and, in specific scenarios, provide higher bandwidth. In this work, we focus on mesh-based networks, as they are implementation-friendly and are the most widely adopted topology in practice \cite{taylor2003scalar,gratz2006implementation}. Each core is directly connected to a router and consists of three major components: a control unit, a compute unit, and a scratchpad memory. The control unit manages intra-core resource allocation and task scheduling. The compute unit executes DNN operators and typically integrates specialized accelerators such as vector units or tensor units optimized for matrix operations. The scratchpad memory is a small, high-speed SRAM placed adjacent to the compute unit, typically less than \SI{64}{\kibi\byte} in capacity, storing activations, weights, and intermediate results. It is explicitly managed by software or compilers, thereby simplifying hardware coherence logic \cite{10.1145/1176887.1176933,banakar2002scratchpad}.

\begin{figure*}[t]
  \centering
  \includegraphics[width=\textwidth]{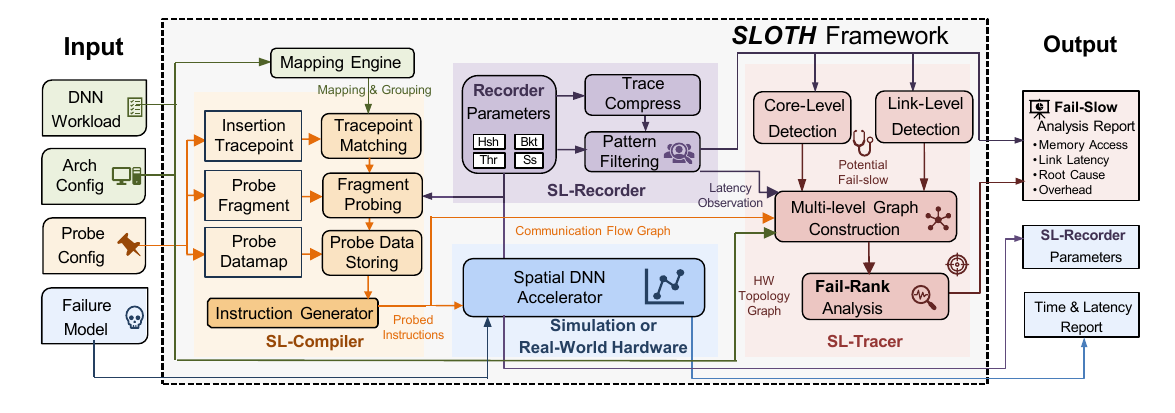}
  \captionsetup{skip=-5pt}
  \caption{Detailed workflow of the proposed \project framework. The system integrates three key components to achieve end-to-end fail-slow detection: (1) SL-Compiler performs workload-aware static analysis to insert lightweight probing fragments at semantic operator boundaries; (2) SL-Recorder operates on-the-fly within each core, employing a two-stage sketch structure to dynamically filter redundant execution patterns and compress traces into SRAM-friendly memory footprints; (3) SL-Tracer constructs a Multi-level Communication Graph (MCG) by fusing software dataflow with hardware topology, and executes the FailRank algorithm to output a ranked list of the most probable fail-slow root causes (cores or links).}
  \label{fig:overview}
  \vspace{-5pt}
\end{figure*}

Based on the hardware organization above, spatial DNN accelerators employ a dataflow execution model to maximize concurrency and overlap computation with communication\cite{chen2016eyeriss,hegde2021mind,cai2024gemini}. In this paradigm, DNN models are represented as computation graphs, where nodes denote operators (e.g., convolution, pooling) and edges represent data dependencies between layers\cite{dennis1974preliminary,nikhil2002executing}. Execution is data-driven---an operator is triggered as soon as its input features and weights become available, avoiding the cost of global synchronization. DNN operators are mapped onto cores according to data locality and the layer-wise computational workflow, while the scheduler within each core monitors operator readiness, fetches operands from local or neighboring cores, and dispatches them to compute units. Intermediate features are routed through the NoC according to the communication pattern derived from the DNN computation graph. This fine-grained pipelining of data movement and operator processing improves hardware utilization, but also creates tight coupling between cores: delays in one core's execution can propagate through dependent operators across the chip, amplifying the performance impact of fail-slow failures in DNN inference.

\subsection{Hardware Failures}

In spatial DNN accelerators, hardware components are subject to a wide range of failures that invariably degrade inference performance. Based on their behavioral characteristics, common hardware failures are classified into two categories: fail-slow and fail-stop\cite{gunawi2018fail,panda2019iaso}. The causes of fail-slow are complex, spatially non-uniform, and span multiple layers of the system.

\textbf{Core-level Failures.} For cores, severe bit flips that alter register values or permanent structural defects typically lead to program crashes \cite{sastry2009mswat,he2023understanding}, which are characteristic of fail-stop failures. Conversely, core-level fail-slow failures arise primarily from physical degradation and thermal stress, exhibiting strong spatial variance across the chip. First, device-level aging \cite{bhardwaj2012towards, haghbayan2020thermal, wang2019optimized}, such as Negative Bias Temperature Instability (NBTI), causes cores to age at different rates depending on their workload history, increasing path delays by 10\% to 40\% \cite{agarwal2007circuit}. Second, sustained high-utilization workloads generate localized thermal hotspots \cite{lin2017thermal,shukla2019overview}. Thermal throttling at these specific regions (e.g., exceeding 80°C in 3D-stacked chips) can degrade local core performance by 24\% to 52\% \cite{pandey20243d}. Over time, these spatially localized slowdowns accumulate, turning individual cores into performance bottlenecks \cite{tang2018vsensor,zheng2022vapro}.

\textbf{Interconnect and Memory Failures.} Beyond cores, the on-chip data delivery infrastructure is also vulnerable. For memory, while permanent physical defects cause immediate crashes, extreme temperatures or buffer contention can trigger protective states that temporarily restrict effective memory bandwidth \cite{lin2017thermal, lu2022nvme}, leading to fail-slow failures. More critically for spatial architectures, the NoC routinely suffers from transient fail-slow degradation. While routing-induced deadlocks halt communication outright (fail-stop) \cite{taheri2022deft}, the spatial mapping of DNNs often forces cores to compete heavily for NoC routing resources. This shared resource contention, coupled with localized voltage droops, can severely diminish link bandwidth and prolong multi-hop transmission times \cite{chang2011design,tsoutsouras2017softrm}.

In this paper, we focus on detecting {fail-slow} failures caused by core-level performance variance and link-level bandwidth degradation—both of which are prevalent and, more importantly, can propagate along communication and execution dependencies, leading to system slowdowns.

\section{The \project Framework}

\project is an automated framework for detecting and locating fail-slow failures in spatial DNN accelerators. As shown in Figure \ref{fig:overview}, the framework is composed of three main components. \textbf{SL-Compiler} (\Cref{sec:sl-compiler}) instruments DNN operators with lightweight probes at operator boundaries, collecting fine-grained runtime execution traces with minimal overhead.
\textbf{SL-Recorder} (\Cref{sec:sl-recorder}) then filters irrelevant traces and compresses redundant ones, producing a compact yet representative list of communication and computation patterns.
Finally, \textbf{SL-Tracer} (\Cref{sec:sl-tracer}) leverages the compressed data to construct multi-level communication graphs and performs the FailRank algorithm on them, pinpointing fail-slow cores or links responsible for performance degradation. Together, these components form a systematic pipeline that bridges low-overhead monitoring with high-accuracy failure diagnosis.

% an overall view of the framework
\subsection{Input and Output}
\label{sec:input-output}

As shown in the left of Figure~\ref{fig:overview}, the inputs to \project consist of: (1) \textbf{DNN Workload}: DNN models are composed of operators such as convolution, pooling, and data movement instructions including \textit{read}, \textit{send}, and \textit{recv}. The framework supports arbitrary operator-to-core mappings~\cite{280874,9076333,hanson2023si}. In our implementation, we use a Gemini-like~\cite{cai2024gemini} mapping strategy for evaluation; (2) \textbf{Arch Config}: the specification of the target spatial DNN accelerator architecture, which enables analysis across different hardware platforms; (3) \textbf{Probe Config}: parameters that determine what execution information to collect, where to place probes at operator boundaries, and how to store the collected trace data; (4) \textbf{Failure Model}: definitions of fail-slow behavior, including failure duration, performance degradation rate, and affected components.

The outputs fall into two categories: \textbf{fail-slow detection results}, including the root causes, SL-Recorder's optimal compression parameters, and the overhead of data collection and storage; and \textbf{system-level metrics}, such as total inference time and per-layer execution latency.

\subsection{SL-Compiler}
\label{sec:sl-compiler}

Traditional tracing approaches face a trade-off between coverage and overhead: exhaustive collection is expensive, while reactive collection risks missing critical failure signatures. To address this challenge, we propose SL-Compiler, a dataflow-graph-guided instrumentation tool. Instead of relying on expensive, interrupt-driven OS monitors, SL-Compiler analyzes the DNN operators of the computation graph at compile-time. It strategically place timing probes exactly at operator boundaries (e.g., between a Conv2D and a Pooling layer). This workload-aware placement avoids disrupting the tight inner-loop pipeline of the tensor compute units, thereby bounding the monitoring overhead to less than 10\% while preserving high-fidelity fail-slow signatures.

\noindent\textbf{Probe Abstraction}. SL-Compiler decomposes the probing into three configurable abstractions, which is formalized as a five-tuple configuration: \texttt{<Fragment,Type,Location, Level,Structure>}. \texttt{Fragment} defines what data to collect, \texttt{Type,Location} specify where to place the probes, and \texttt{Level,Structure} determine how to organize the collected traces. SL-Compiler can recognize \texttt{Fragment} and \texttt{Type} automatically, while the remaining three parameters are user-configurable. Users can express complex monitoring strategies according to their requirements. If not configured, the default value is set \texttt{<Location=Surround,Level=Stage, Structure=Sketch>}, the same settings used in evaluation.

\begin{figure}[tbp]
    \centering
    \includegraphics[width=\linewidth]{probe-new-drawio.pdf}
    \captionsetup{skip=3pt}
    \caption{Example of SL-Compiler probing. By analyzing the computation graph, the compiler strategically inserts probes. After compute op analysis, SL-Compiler inserts \texttt{<Exec,Comp,Post,Stage,List>} probe. After communication op analysis, SL-Compiler inserts \texttt{<Route,Comm,Pre,Inst,Sketch>} probe.}
    \label{fig:probe-new}
    \vspace{-10pt}
\end{figure}

\noindent\textbf{DNN-graph-guided Insertion.} A key innovation of SL-Compiler is its ability to automatically generate probe configurations by analyzing the DNN computation graph. It parses the computation graph to extract the layer sequence, data dependencies, and operator types (e.g., Conv2D, Pooling), and then chooses suitable probe fragment according to operator type. All the operators are instrumented during the trace collection, while off-chip operators (memory-intensive ops like Read and Write) are excluded in detection phase since they are bottlenecked by memory controllers and DRAM bandwidth, which mask the intra-chip performance variations that SLOTH targets. Different probe configurations can coexist when SL-Compiler deems both necessary. As shown in Figure \ref{fig:probe-new}, SL-Compiler analyzes compute operator first and decides to insert post-execution probe \texttt{<Exec,Comp,Post,Stage,List>}. Then SL-Compiler checks communication operators and finds it necessary to insert pre-communication probe \texttt{<Route,Comm,Pre,Inst,Sketch>} to monitor link latency. Finally, it generates the instrumented code by inserting lightweight probe fragments at the tracepoints identified by the five-tuple configurations. Additionally, to deal with the data-dependent control flow in dynamic DNNs, SL-Compiler adds a lightweight control-flow marker probe at the end of each branch. The probe returns a specific Branch\_ID value during runtime, which is then parsed by SL-Tracer to construct exact MCG.

\begin{table}[tbp]
\vspace*{1mm}
% \footnotesize
\centering
\captionsetup{skip=3pt}
\caption{Components of probe's five-tuple definition.}
\label{Compiler-API}
\begin{tabular}{ccl}
\toprule
\textbf{Component} & \textbf{Option} & \textbf{Description} \\
\midrule
\multirow{3}{*}{\texttt{Fragment}}
  & \texttt{Exec} & Record calculation related traces \\
  & \texttt{Route} & Record communication traces \\
  & \texttt{Mem} & Record memory access traces \\
\midrule
\multirow{3}{*}{\texttt{Type}}
  & \texttt{Comp} & Match compute instructions \\
  & \texttt{Comm} & Match communication instructions \\
  & \texttt{IO} & Match input / output instructions \\
\midrule
\multirow{3}{*}{\texttt{Location}}
  & \texttt{Pre} & Before the target instruction \\
  & \texttt{Post} & After the target instruction \\
  & \texttt{Surround} & Before \& after the instruction \\
\midrule
\multirow{2}{*}{\texttt{Level}}
  & \texttt{Inst} & Aggregate trace at instruction level \\
  & \texttt{Stage} & Aggregate trace at stage level \\
\midrule
\multirow{2}{*}{\texttt{Structure}}
  & \texttt{List} & Store traces in list manner \\
  & \texttt{Sketch} & Store traces using SL-Recorder \\
\bottomrule
\end{tabular}
\vspace{-18pt}
\end{table}

\noindent\textbf{Operator-level Semantic Collection}. Beyond basic performance metrics, SL-Compiler captures high-level DNN operator semantics to facilitate root cause analysis. Each probe records not only the standard execution attributes---such as start/end timestamps, and core ID, but also DNN-specific metadata that characterizes the computational context. For computation probes, this includes the operator type (e.g., Conv2D, GEMM), input/output tensor dimensions, and floating-point operation count (FLOPs). For communication probes, it records the data volume transferred, the source and destination cores, and the corresponding tensor being communicated. This semantic information serves two critical purposes: (1) it enables layer-aware analysis in SL-Tracer, allowing the framework to correlate slowdowns with specific DNN layers; and (2) it facilitates communication pattern recognition, helping distinguish between normal inter-layer data transfers and anomalous traffic caused by fail-slow failures. For example, if a Conv layer exhibits unexpectedly high communication latency, the recorded tensor dimensions can reveal whether the slowdown stems from excessive feature map sizes or from degraded link bandwidth.

% \vspace{-15pt}

\subsection{SL-Recorder}
\label{sec:sl-recorder}

\begin{figure}[tb]
  \centering
  \includegraphics[width=\linewidth]{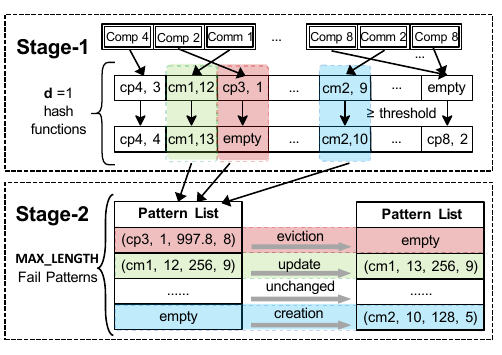}
  \captionsetup{skip=0pt}
  \caption{An example of Fail-Slow Sketch (with a single hash function, d=1). In Stage-1, incoming traces are mapped into hash buckets to identify frequent workload patterns. Stage-2 maintains a bounded list that continuously updates statistical summaries of compressed traces.}
  \label{fig:sketch}
  \vspace{-10pt}
\end{figure}

To reduce the storage overhead, we propose SL-Recorder, a lightweight \textbf{online data structure} for trace compression and pattern-based filtering. The data structure is also called Fail-Slow Sketch as it is inspired by sketches.
Importantly, SL-Recorder requires no hardware modifications; the probe instructions are inserted statically by SL-Compiler at compile time, so the only costs are runtime and storage, both of which we quantify in \Cref{sec:overhead}.
As illustrated in Figure \ref{fig:sketch}, Fail-Slow Sketch operates in two stages. In Stage-1, a Running Track identifies frequently occurring traces with similar workload pattern. In Stage-2, the steady-state redundancy of the traces are extracted to reduce storage while keeping deviations of fail-slow onset.

\SetKwInput{SkeInput}{Input}
\SetKwFunction{SkeStage}{Insert\_Stage\_2}
\SetKwFunction{SkeUpd}{Update}
\SetKwProg{Fn}{Function}{}{}

\begin{algorithm}[tbp]
% \footnotesize
\caption{Fail-Slow Sketch Insertion}
\label{algo:insertion_modified}
\SkeInput{instruction trace $e$; number of hash tables $d$; number of buckets $m$; pattern threshold $H$}
\For{each $i \in [1, d]$}{
    $Index \gets \text{Hash}(e.key) \bmod m$\;
    $bucket \gets \text{HashTable}[i][Index]$\;

    \If{bucket.key equals to e.key}{
        increase the frequency of e by 1\;
        % --- MODIFICATION 1: Use \lIf for single-line if ---
        \lIf{frequency of $e \geq H$}{
            \SkeStage{e, attributes}
        }
    }
    \ElseIf{bucket is empty}{
        $bucket.key \gets e.key$\;
        set the frequency of e to 1\;
    }
    \Else{
        decrease the frequency of e by 1\;
        % --- MODIFICATION 2: Use \lIf for single-line if ---
        \lIf{frequency of e is 0}{
            clear $bucket$ to empty\
        }
    }
}

\vspace{0.5em}
\Fn{\SkeStage{$e$, attr}}{
    \If{$e.key \notin \text{Stage-2}$}{
        % --- MODIFICATION 3: Use \lIf for single-line if ---
        \lIf{Stage-2.length $\geq$ \texttt{MAX\_LENGTH}}{
            evict the earliest pattern in Stage-2\
        }
        insert $e$ into Stage-2\;
        \KwRet 1\;
    }
    \Else{
        \SkeUpd{$e.key$, attr}\;
    }
    \KwRet 0\;
}
% \vspace{-8pt}
\end{algorithm}

\textbf{Stage-1} consists of $d$ bucket arrays, each associated with a distinct hash function. These hash functions map trace features that characterize execution patterns to bucket indices. Each array contains $m$ buckets, and each bucket has two fields: trace pattern (key) and its frequency. When the frequency exceeds a threshold $H$, the corresponding trace pattern is considered as a potential candidate for fail-slow pattern.

\textbf{Stage-2} maintains a list of candidate fail-slow patterns and their associated attributes. This list supports operations such as updates and evictions, and its maximum length (\texttt{MAX\_LENGTH}) is determined based on the characteristics of the workload. In this stage, traces with the same pattern are compressed, and statistical summaries are computed to support root cause analysis. When the list reaches its maximum capacity, the Fail-Slow Sketch evict the most ``healthy'' pattern to make room for newly observed patterns.

\textbf{Insertion}. Given an input trace $t$, Fail-Slow Sketch first extracts its pattern features and computes the target bucket index using corresponding hash function. The trace is then inserted into the bucket (donated as $b$). As shown in \Cref{algo:insertion_modified}, there are three possible cases:

\textit{Case 1:} $t$ is not in the bucket, and $b$ is currently empty. In this case, the trace pattern is stored in $b$, the frequency is set to 1, and the insertion completes.

\textit{Case 2:} $t$ is not in the bucket, and $b$ is occupied. In this case, the frequency of $b$ is decremented by 1. If the frequency drops to zero, the bucket is cleared, and the algorithm terminates.

\textit{Case 3:} $t$ is in the bucket $b$. In this case, the frequency counter is incremented by 1. If the frequency reaches the threshold $H$, the trace pattern is considered a candidate for Stage-2. If the pattern already exists in the Stage-2 list, its statistics are updated via update function; otherwise, a new pattern entry will be created. If the Stage-2 list has reached its capacity, typical replacement strategies such as frequency-based, probability-based, or arrival-time-based policies can be applied. In our design, we use an arrival-time-based replacement policy to ensure correct insertion of trace patterns.

% \textbf{Update}. Given a trace pattern and its attributes, the update operation modifies the corresponding entry in the Stage-2 pattern list to maintain the overall statistical information. For communication patterns, this includes updating metrics like data volume. For computation patterns, the update includes floating-point operation count, core id, and other relevant metrics. Additionally, all trace patterns have to maintain the timestamp of their first appearance and their duration, which are essential for root cause analysis.

\textbf{Update}. Given a trace pattern and its attributes, the update operation modifies the corresponding entry in the Stage-2 pattern list to maintain the overall statistical information. The pattern matcher classifies each trace entry using two features: (1) \textbf{spatial location} (executing/src/dst core ID) and (2) \textbf{operator resource footprint} (memory consumption, FLOPs, etc.). Entries matching a previously observed pattern are compressed in-place, reducing storage and downstream analysis costs without requiring offline post-processing. For communication patterns, this includes updating metrics like data shapes. For computation patterns, the update includes average FLOPs and other relevant metrics. Additionally, all trace patterns maintain the timestamp of their first appearance and their duration, which are essential for root cause analysis.

\textbf{Example}: Figure \ref{fig:sketch} shows a running example of Fail-Slow Sketch. To insert coming trace $Comp4$, Stage-1 extracts its pattern $cp4$. $cp4$ has been inserted into corresponding bucket before, so Fail-Slow Sketch adds the frequency of the bucket by 1 and finishes inserting. When inserting trace $Comm2$, the frequency of the bucket reaches threshold $H = 10$ and there is space in Stage-2's pattern list. Therefore, Fail-Slow Sketch create a new trace pattern ($cm2$,10,128,5) for $Comm2$ through \textit{Creation} operation, in which $cm2$ is the trace pattern, 10 is $cm2$'s frequency, 128 is the data volume and 5 is the destination of communication. Besides, Fail-Slow Sketch performs \textit{Update} operation for existing pattern $cm1$ and \textit{Eviction} operation for deleted pattern $cp3$. After inserting all the traces, the compressed traces, also the trace patterns are stored in Stage-2's pattern list.

\textbf{Mathematical Analysis}. It can be proved that increasing the number of hash functions $d$ can effectively improve the retention probability, but also increases computational overhead; increasing the number of buckets $m$ helps reduce hash collisions, but also increases storage overhead. We provide a proof for the conclusion as follow.

Given a fixed workload, let $N$ be the number of instruction traces, $f_i$ be the frequency of trace $t_i$, $B_j[k]$ be $k$-th bucket of $j$-th bucket array, and $F_{i,j,k}$ be the number of traces mapped to bucket $B_j[k]$ except for trace $t_i$.

\textsc{LEMMA 3.1.} For any trace pattern $R_i$ in the instruction traces, let $P(R_i)$ be its retention probability in Stage-2, then
$$ P(R_i) \ge 1 - (\frac{N-f_i}{m(f_i-H)})^d $$

\textsc{PROOF.} For a single bucket $A_j[k]$ in Stage-1 which records trace $t_i$, according to the insertion algorithm, we have
$$ f_i - F_{i,j,k} \le A_j[k].freq \le f_i$$
Since the hash function is uniformly distributed, the probability of mapping to a specific bucket is $\frac{1}{m}$, so the expectation of $F_{i,j,k}$ is
$$ E[F_{i,j,k}] = \Sigma_{t\ne t_i} f_t \frac{1}{m} = \frac{N-f_i}{m} $$
For a trace pattern $R_i$ in Stage-2 and threshold $H$, there must be a $j \in \{1,2,...,d\}$ such that
$$ f_i-F_{i,j,k} \ge H \Leftrightarrow F_{i,j,k} \le f_i-H $$
By Markov inequality, we can evaluate the probability of not recoding trace pattern $R_i$ with one hash function
$$ P(F_{i,j,k} \ge f_i-H) \le \frac{E[F_{i,j,k}]}{f_i-H} = \frac{N-f_i}{m(f_i-H)} $$
Since we have $d$ hash functions, so the retention probability of a trace pattern $R_i$ is
$$ P(R_i) \ge 1 - (\frac{N-f_i}{m(f_i-H)})^d $$

We further analyze compression parameter sensitivity and the accuracy--overhead tradeoff to determine the optimal SL-Recorder configuration (\Cref{sec:overhead}).

\subsection{SL-Tracer}
\label{sec:sl-tracer}

SL-Tracer is designed to identify and localize the root cause fail-slow failures. It follows a detection–construction–analysis pipeline. First, core-level detection and link-level detection are performed to identify potential fail-slow candidates. These candidates, together with workload’s communication dependency graph and the hardware topology, are integrated together to construct multi-level communication graph (MCG). Finally, the FailRank algorithm operates on this graph to quantify the likelihood of each candidate being the root cause, providing a ranked explanation of fail-slow components.

\subsubsection{Core-Level Fail-Slow Detection}
\label{sec:core-level}

The goal of core-level fail-slow detection is to identify candidate fail-slow cores. Our method performs the detection in a stage-aware manner, leveraging the mapping from DNN computation graph to spatial accelerator.

\begin{figure}[t]
  \centering
  \includegraphics[width=\linewidth]{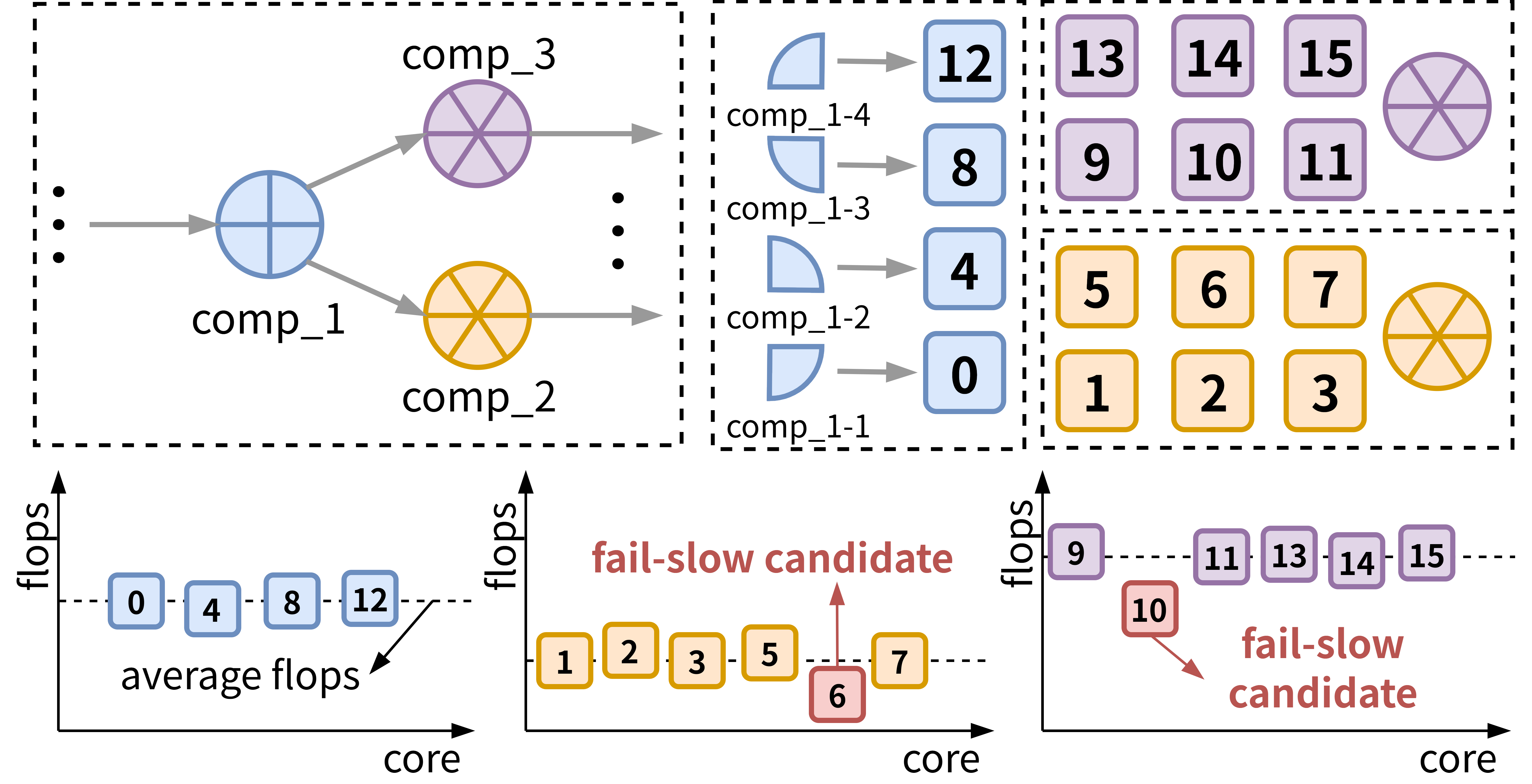}
  \captionsetup{skip=3pt}
  \caption{Core level fail-slow detection. Computation instructions are first partitioned by execution stages and clustered according to the workload-to-core mapping. By comparing the processing throughput (FLOPs per unit time) against the intra-group baseline, anomalous outlier cores (e.g., Core 6 and 10) are flagged as fail-slow candidates and assigned initial anomaly probabilities.}
  \label{fig:core-level-v2}
  \vspace{-25pt}
\end{figure}

As illustrated in Figure \ref{fig:core-level-v2}, we first partition computation instructions by execution stage (e.g., network layers), ensuring that only tasks with comparable semantics are grouped together. Next, cores are clustered according to the workload-to-core mapping. Each group therefore contains volume-equivalent computations executed across different cores, making their performance directly comparable.

Within each group, we measure the performance of each core in terms of floating-point operations (FLOPs) per unit time. The group average serves as the baseline, and we perform outlier detection to highlight cores whose FLOPs are significantly lower than the baseline. These outliers are flagged as fail-slow candidates. For example, in Figure \ref{fig:core-level-v2}, core 6 and 10 deviate from their peers and are thus marked as candidates. Each candidate is further assigned a fail-slow probability, derived from its performance variance distribution. This probability is then used as the initial fail-slow score for the FailRank algorithm (Sec \ref{sec:RCA}).

\subsubsection{Link-Level Fail-Slow Detection}
\label{sec:link-level}

The goal of link-level fail-slow detection is to identify candidate fail-slow interconnect links. Since such fail-slow behaviors cannot be directly observed under limited hardware visibility, we infer them from communication traces collected by SL-Recorder. Specifically, we formulate the problem as a statistical inference task, where observed end-to-end communication delays are decomposed to estimate the bandwidth of individual links.

For each communication event, we record four key attributes: the transferred data volume, the measured transmission time, and the index of the source and destination cores. Leveraging the deterministic routing algorithm of the 2D-mesh interconnect, we map each event to the set of links traversed along its multi-hop path, thereby associating observed end-to-end latency with individual link bandwidths.

To estimate link bandwidth, we transform the problem into a set of linear equations by turning each link's bandwidth $b_i$ into its inverse $\displaystyle \theta_i = \frac{1}{b_i}$. Each communication event yields an equation of the form $Ax^{T} = T_{comm}$, where $A$ represents the link composition of paths, $x$ represents $\theta_i$ vector and $T_{comm}$ represents the observed transmission times.

Because the system is underdetermined and affected by measurement noise, we adopt the Expectation–Maximization (EM) algorithm to infer link bandwidths. EM algorithm iteratively estimates link bandwidths by alternating between (1) computing expected path-level delays given current bandwidth estimates and (2) updating bandwidth estimates to maximize the likelihood of the observed traces. This process converges to a set of link bandwidths that best explain the data, even under partial or noisy observations.

To assess the likelihood of fail-slow behavior, we calculate the fail-slow probability using the bandwidth gamma distribution model. These probabilities are then assigned as the initial fail-slow scores of links in the FailRank algorithm.

\subsubsection{Fail-Slow Root Cause Analysis}
\label{sec:RCA}

To localize fail-slow root causes, we construct MCG that integrates hardware topology, execution dependencies, and runtime tracing data. This augmented graph captures how performance degradation propagates across cores and links, turning the localization into a graph-based inference problem. On this foundation, we design \textbf{FailRank}, a PageRank-inspired algorithm that quantifies the propagation probability of fail-slows, enabling accurate root cause identification.

\begin{figure}[tbp]
  \centering
  \includegraphics[width=\linewidth]{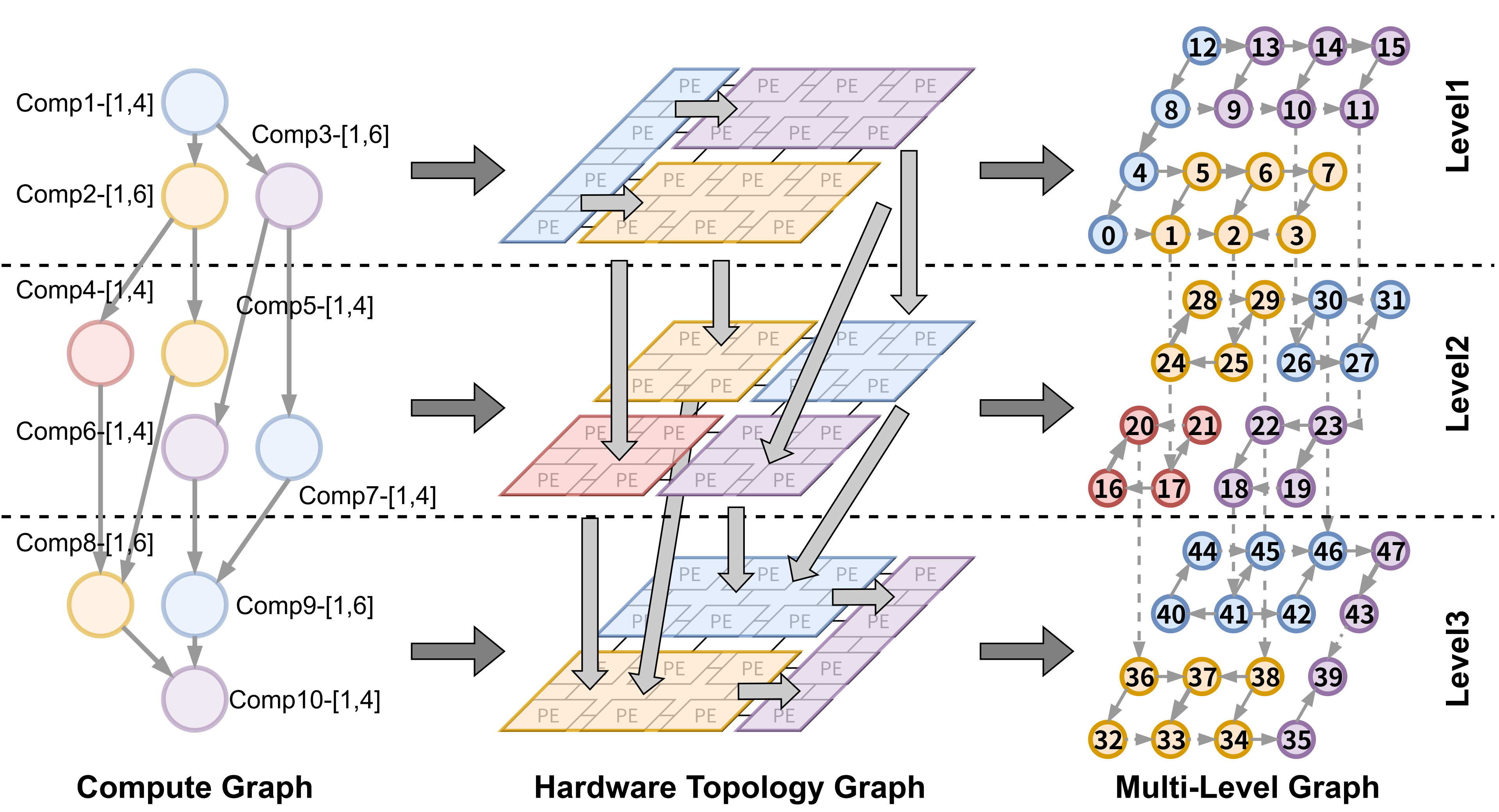}
  \captionsetup{skip=3pt}
  \caption{Construction of multi-level communication graph (MCG). SL-Tracer maps the logical data dependencies of the computation graph (DAG) onto the physical NoC hardware topology graph.}
  \label{fig:multi-level}
  \vspace{-10pt}
\end{figure}

\textbf{MCG Definition.} MCG is defined as a layered graph $G = (V,E)$ that integrates task dependency and hardware topology, shown in the right of Figure \ref{fig:multi-level}. Each node $v \in V$ represents a core in a time window, bounded with a set of computation instructions. The likelihood for being root cause of node $v$ is quantified by a fail-slow score $s(v) \in [0,1]$ (determined in Sec \ref{sec:core-level}). Each directed edge $e=(u,v) \in E$ denotes communication dependency between two cores, bounded with a set of communication instructions. The edge is associated with a propagation weight $w(u,v)$, capturing the probability that performance degradation propagates from core $u$ to $v$.

To construct the MCG, we perform a depth-first search (DFS) on the workload’s computation graph $G_c = (V_c, E_c)$. As illustrated in Figure \ref{fig:multi-level}, $G_c$ is a directed acyclic graph (DAG). For each dependency edge $(x,y) \in E_c$, we first determine the physical cores $p(x)$ and $p(y)$ assigned to tasks $x$ and $y$. The mapping $p:V_c \rightarrow E_h $ links computation nodes to hardware cores in the hardware topology $G_h = (V_h, E_h)$. The communication path between $p(x)$ and $p(y)$ is then identified as a sequence of hardware links $\{e_1, e_2, ..., e_k\} \in E_h$. For each traversed link $e_i$, we increment its traffic volume $T(e_i)$ by the transported data size of $(x,y)$.

After traversing the entire DAG, we normalize the traffic volumes to derive edge propagation weights. Specifically, for each core $u \in V_h$, let $Out(u) = \{(u,v) \in E_h\}$ denote its outgoing links. This normalization ensures that $\Sigma_{(u, v^{'}) \in Out(u)} w(u, v^{'}) = 1$, making $w(u,v)$ interpretable as the probability of failure propagation through link $(u,v)$. The resulting MCG thus provides a structured, probabilistic view of how fail-slow behaviors may spread across both computation and communication dependency within a level.

To capture cross-level interactions, we introduce virtual DRAM nodes to connect levels. These nodes represent interactions  through memory, such as fetches or writes between different levels. Edges connected to a DRAM node represent inter-level dependencies, showing how fail-slow may propagate across levels. This vertical connection adds a temporal dimension to MCG, allowing the FailRank algorithm to analyze the timing and causality of fail-slows.

\textbf{FailRank Algorithm.} To localize fail-slow root causes from core-level and link-level candidates, we propose FailRank, a graph-based iteration algorithm inspired by PageRank, which identifies fail-slow root causes on MCG. Unlike PageRank that only updates node scores based on their neighbors with uniform transition probabilities, FailRank incorporates propagation weights of their neighbor links.

The algorithm initialize fail-slow scores with the probabilities obtained in core-level and link-level detection, denoted as $s_0(v)$ for node $v$ and $l_0(u,v)$ for edge $(u,v)$. At each iteration, a node’s score is updated not only from its neighboring nodes but also from the weighted contributions of connecting links:
$$s^{(k+1)}(v) = (1-\lambda)s_0(v) + \lambda \Sigma_{(u,v) \in E}w(u,v)s^{(k)}(u)$$
where $w(u,v)$ is the normalized propagation weight of edge $(u,v)$ defined in Sec \ref{sec:RCA}. Each link's score is updated by three factors: the propagation weight, the fail-slow scores of its endpoint node $u$, and its current score:
$$l^{(k+1)}(u,v) = \alpha w(u,v) + \beta s^{(k)}(u) + \gamma l^{(k)}(u,v)$$
where $\alpha,\beta,\gamma$ reflect the relative importance of communication versus computation propagation.
Users can flexibly tune the update weights according to workload characteristics.
In our experiments, we fix $\alpha=0.1, \beta=0.3, \gamma=0.6$ across all workloads to demonstrate robustness under a single configuration.

\begin{figure}[tb]
  \centering
  \includegraphics[width=\columnwidth]{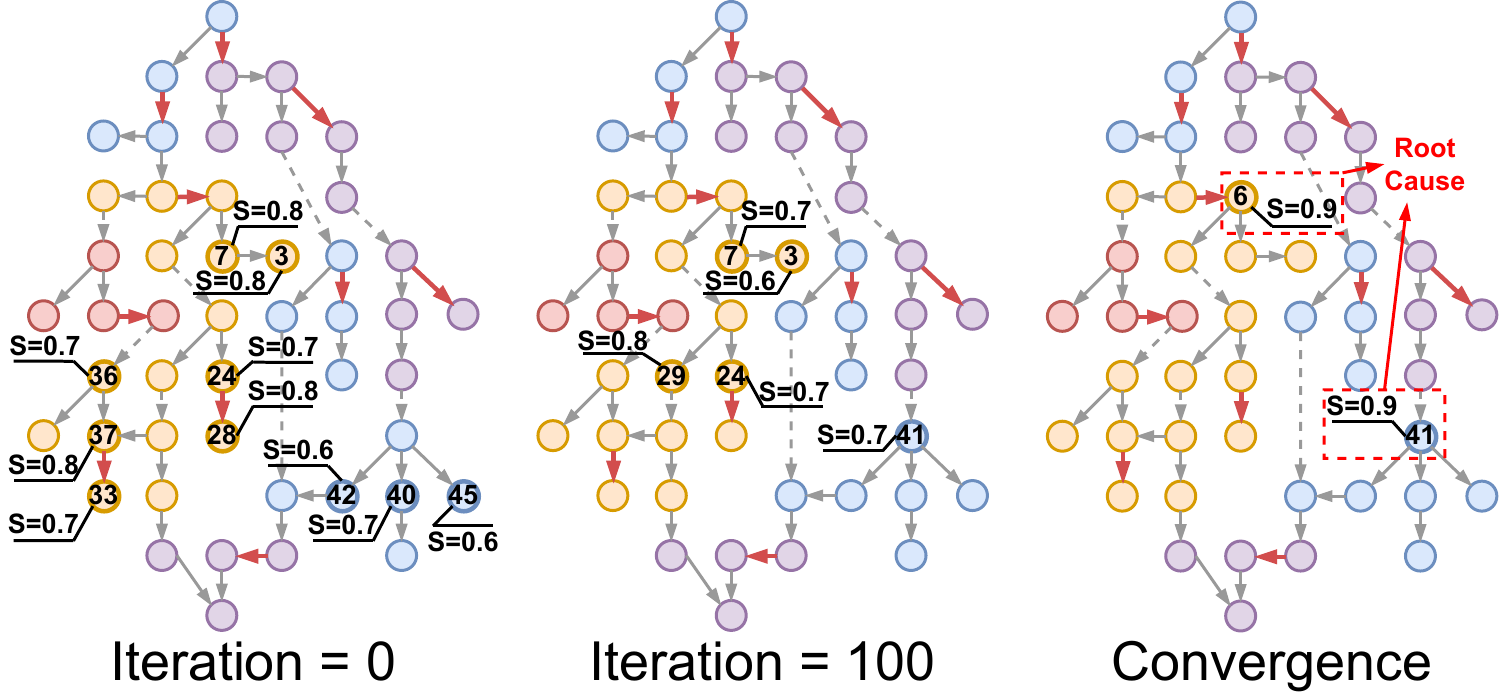}
  % \captionsetup{skip=3pt}
  \caption{An example of FailRank algorithm. Starting from the initial candidate probabilities, FailRank iteratively updates each node's fail-slow score based on its neighbors and the weighted links. Upon convergence, the node with the highest probability is pinpointed as the root cause.}
  \label{fig:failrank}
  \vspace{-18pt}
\end{figure}

As Figure \ref{fig:failrank} shows, FailRank starts with initial fail-slow candidates and iteratively updates their scores until convergence.
Each node is assigned a fail-slow score $s$, as annotated in the figure (e.g., $s=0.8,0.7$), which reflects the likelihood of being the root cause.
Edges indicate communications between cores. Red edge has higher communication volume, while the dashed edge shows memory-access dependencies between levels.
During iterations, the nodes with higher fail-slow scores and edges with heavier traffic exert stronger influence on their neighbors.
After convergence, FailRank normalizes node and edge scores within each level of the MCG using a softmax function, yielding a probability distribution that pinpoints the most likely fail-slow root cause.
\project operates at inference granularity: SL-Recorder updates statistics continuously during execution, and SL-Tracer produces a root-cause ranking at inference completion.
Detection latency is therefore bounded by one inference pass.

\section{Evaluation}

We evaluate \project along four dimensions:
(1)~detection accuracy against five baselines
(\S\ref{sec:detection});
(2)~runtime and storage overhead, including the accuracy--overhead tradeoff
(\S\ref{sec:overhead});
(3)~sensitivity to fail-slow severity
(\S\ref{sec:sensitivity});
and (4)~scalability across accelerator sizes and topologies
(\S\ref{sec:scalability}).

\subsection{Experimental Setup}

\noindent\textbf{Evaluation Platform}. We implement an event-driven cycle-level simulator for spatial DNN accelerators. The simulator models two key components: (1) core microarchitecture, including the task scheduler, a general compute unit, and scratchpad memory; and (2) NoC interconnect, including per-router arbitration, per-hop latency, and link contention under multi-flow traffic. We validate the simulator with the Dataflow Abstract Machine (DAM) framework~\cite{10.1109/ISCA59077.2024.00046} against the same microbenchmarks: binary reduction trees with \{2, 8, 32\} trees, depths of \{8, 10\}, and 100K reductions per tree. The results show less than 10\% deviation across all configurations.

Standard cycle-level simulators for dataflow accelerators produce deterministic execution traces,
under which any injected fail-slow is trivially separable from normal behavior.
To ensure our evaluation reflects realistic detection conditions,
we augment cycle-accurate timing with stochastic hardware variability,
following established modeling methodology~\cite{romanescu2006quantifying,hernandez2010methodology}
and calibrating all parameters to distributions measured on production silicon~\cite{inadomi2015analyzing,marathe2017empirical}.
This noise floor forces the detector to distinguish genuine fail-slow deviations from natural performance spread,
substantially raising detection difficulty. Concretely, each core's computing capacity is modeled as $P_{core} \sim N(\mu_c, \sigma^2_c)$, where $\mu_c$ is the expected throughput and $\sigma^2_c$ captures runtime variability.
Link transmission time follows $T_{link} \sim \text{Gamma}(\alpha, \beta)$, capturing the right-skewed latency distribution arising from contention and bandwidth fluctuation.

\begin{table*}[tbp]
\centering
\caption{Fail-slow detection accuracy and False Positive Rate (FPR).}
\label{fig:detection-acc}
\resizebox{\textwidth}{!}{%
\begin{tabular}{@{} l rrrrrrr @{}} % 1列左对齐(算法名)，7列右对齐(数据)
\toprule
\textbf{Algorithm} 
& \textbf{DarkNet-19} 
& \textbf{GoogLeNet} 
& \textbf{VGG} 
& \textbf{ResNet-50} 
& \textbf{Binary Tree} 
& \textbf{Early-Exit ResNet} % <-- 请替换为你的新Workload 1
& \textbf{Qwen2.5-0.5B} % <-- 请替换为你的新Workload 2
\\
\midrule
% ==================== 第一部分: Accuracy ====================
\multicolumn{8}{c}{\textbf{Detection Accuracy $\uparrow$ (\%)}} \\
\midrule
Z-Thres 
& 4.87(798/16399) 
& 4.87(798/16399) 
& 4.88(798/16343) 
& 4.87(798/16399) 
& 4.95(798/16120) 
& 4.88(798/16355) % <-- 新W1 Acc占位
& 1.13 (800/70798) % <-- 新W2 Acc占位
\\
Mscope  
& 50.00(541/1082) 
& 50.23(551/1097) 
& 48.41(532/1099) 
& 49.40(536/1085) 
& 58.62(554/945) 
& 49.58(538/1085) 
& 34.90(517/1481) 
\\
IASO    
& 27.39(429/1566) 
& 34.79(413/1187) 
& 34.94(414/1185) 
& 11.61(614/5287) 
& 44.26(540/1220) 
& 37.94(409/1078)
& 21.68 (412/1900)
\\
Perseus 
& 57.44(679/1182) 
& 55.46(676/1219) 
& \textbf{82.20(679/826)} 
& 13.02(679/5214) 
& 60.10(693/1153) 
& 33.42(676/2023)
& 37.16 (401/1079)
\\
ADR     
& 9.81(679/6918) 
& 9.54(679/7120) 
& 8.99(679/7557) 
& 8.91(679/7622) 
& 11.24(687/6110) 
& 9.84(679/6902)
& 2.60 (680/26120)
\\
\project (Ours) 
& \textbf{74.59(678/909)} 
& \textbf{75.33(693/920)} 
& 81.82(675/825) 
& \textbf{74.86(667/891)} 
& \textbf{86.69(710/819)} 
& \textbf{81.24(680/837)} % <-- 新W1 Acc Ours
& \textbf{69.68(665/940)} % <-- 新W2 Acc Ours
\\
\midrule
% ==================== 第二部分: FPR ====================
\multicolumn{8}{c}{\textbf{False Positive Rate (FPR) $\downarrow$ (\%)}} \\
\midrule
Z-Thres 
& 97.51(15601/16000) 
& 97.51(15601/16000) 
& 97.50(15545/15944) 
& 97.51(15601/16000) 
& 97.46(15322/15721) 
& 97.50(15557/15956) % <-- 新W1 FPR占位
& 99.43 (69998/70398) % <-- 新W2 FPR占位
\\
Mscope  
& 41.58(284/683) 
& 42.84(299/698) 
& 43.00(301/700) 
& 41.84(287/686) 
& 26.92(147/546) 
& 41.92(288/687)
& 63.12(683/1082) 
\\
IASO    
& 65.81(768/1167) 
& 49.37(389/788) 
& 49.24(387/786) 
& 91.84(4489/4888) 
& 51.40(422/821) 
& 41.32(281/680)
& 73.33 (1100/1500)
\\
Perseus 
& 49.04(384/783) 
& 51.34(421/820) 
& 6.56(28/427) 
& 91.71(4416/4815) 
& 47.08(355/754) 
& 75.45(1226/1625)
& 41.09 (279/679)
\\
ADR     
& 93.88(6120/6519) 
& 94.06(6322/6721) 
& 94.43(6759/7158) 
& 94.48(6824/7223) 
& 93.01(5312/5711) 
& 93.86(6104/6503)
& 98.44 (25320/25720)
\\
\project (Ours) 
& \textbf{21.76(111/510)} 
& \textbf{23.42(122/521)} 
& \textbf{6.34(27/426)} 
& \textbf{18.90(93/492)} 
& \textbf{5.00(21/420)} 
& \textbf{8.90(39/438)} % <-- 新W1 FPR Ours
& \textbf{26.24(142/541)} % <-- 新W2 FPR Ours
\\
\bottomrule
\end{tabular}
} % \resizebox结束
\end{table*}

\noindent\textbf{Workloads.}
We adopt the same DNN workloads and mapping configurations as
Gemini~\cite{cai2024gemini}, a widely adopted open-source DNN accelerator
framework:
GoogLeNet~\cite{szegedy2015going} and
DarkNet-19~\cite{8100173} (parallel branching, complex
synchronization),
VGG~\cite{simonyan2015deepconvolutionalnetworkslargescale}
(deep linear pipeline, memory-intensive), and
ResNet-50~\cite{7780459} (residual connections, prevalent in modern
DNNs).
In addition to the workloads available in Gemini, we evaluate on Early-Exit ResNet~\cite{7900006} and Qwen2.5-0.5B~\cite{hui2024qwen25codertechnicalreport}.
% Together these cover computation-bound, communication-bound, and
% mixed workload scenarios.
For controlled analysis, we construct the same synthetic benchmark as in DAM\cite{10.1109/ISCA59077.2024.00046}. It is a binary-tree computation of matrix operations isolating
communication patterns from DNN-specific semantics.
All workloads are derived from real-world DNN workloads and all execution traces are collected from our cycle-accurate simulator.

\noindent\textbf{Fail-Slow Dataset.} To evaluate detection robustness under realistic hardware failure scenarios, we construct a comprehensive dataset characterized by three key parameters: a fixed slowdown rate of 10$\times$, a duration uniformly distributed
% between 0–10 s
, and a location type (core-level or link-level).
 % at a 7:3 ratio).
Given that fail-slow failures are rare events relative to hard errors, the probability of two independent fail-slow events overlapping within a single inference window is vanishingly small;
we therefore adopt the standard single-fault assumption used throughout the fault-tolerance literature~\cite{Srinivasan2003RAMPA,panda2019iaso}.

(1) \textbf{Spatial-Pattern-Based Dataset.} The first category models static defects stemming from manufacturing process variations and lithography imperfections. We reference real-world wafer map failure patterns\cite{wu2014wafer}-specifically Center, Edge-Ring, Scratch, and Random-to simulate spatial correlations of on-chip defects. For each pattern, we project the pattern onto the accelerator's core array to generate a spatial probability distribution, where cores in specific geometric regions produce higher failure probabilities. Based on this distribution, we sample 100 independent failure cases for each pattern. This probabilistic approach ensures that the dataset reflects realistic spatial distributions of potential weak cells.

(2) \textbf{Workload-Thermal-Based Dataset.} The second category captures dynamic, runtime failures. Extensive system-level reliability studies\cite{Srinivasan2003RAMPA,gunawi2018fail} have established that sustained high-utilization workloads lead to localized thermal hotspots. So we simulate heat accumulation by profiling the workload utilization of each core. A workload-dependent probability map is calculated for each DNN execution trace, where cores with higher utilization intensity are assigned higher failure probabilities. Similar to the spatial dataset, we generate failure cases by injecting failures into these high-risk regions. This models the physical reality where hotter cores are more prone to timing violations and thermal throttling, providing a realistic testbed for workload-aware detection.

\noindent\textbf{Baselines.} No prior work addresses on-chip fail-slow detection directly.
We construct two classes of baselines:
one native on-chip statistical method, and four distributed-system detectors adapted to the on-chip setting.
All methods share the same raw trace collection for fair comparison,
and we perform a full parameter search on our dataset for each baseline.
For methods providing risk-level outputs, we classify
moderate risk or higher as fail-slow.

(1) \textbf{Z-score Filtering (Z-Thres)} is the natural on-chip baseline:
each component independently monitors its own hardware performance metrics by maintaining running statistics,
and flags itself when its Z-score exceeds 3 (the 3$\sigma$ rule).
We choose it over fixed-threshold or moving-average detector because
Z-Thres can adapt to each component's individual performance distribution.

(2) \textbf{Microscope (Mscope)~\cite{10.1007/978-3-030-03596-9_1}} constructs a directed acyclic graph of inter-service dependencies and ranks root causes using random walk-based scoring. We model data dependencies between operators as service dependencies.

(3) \textbf{IASO~\cite{panda2019iaso}}. A peer-based detection framework that converts timeout signals into scores using AIMD combined with DBSCAN clustering. We treat each core as a peer node, using execution time violations as timeout signals.

(4) \textbf{Perseus~\cite{10.5555/3585938.3585942}} builds polynomial regression models on latency-versus-throughput distributions, employing PCA and DBSCAN for outlier detection. We construct core-level model and use the 99.9th percentile as the detection threshold.

(5) \textbf{Adaptive Detection at Runtime (ADR)~\cite{10.5555/3767955.3767975}} monitors both value and update frequency of performance variables using sliding windows with adaptive thresholds based on historical percentiles.

\subsection{Fail-slow Root Cause Detection}
\label{sec:detection}

Table~\ref{fig:detection-acc} summarizes detection accuracy and false
positive rate (FPR; mis-identification rather than missed detection) across seven workloads.

Baseline methods are limited by missing either detection scope or
topology awareness.
Z-Thres and Mscope detect both core and link faults but suffer from low
sensitivity (Z-Thres misses over 95\% of true faults) and high FPR
(Mscope reaches 26-63\%).
IASO, Perseus, and ADR achieve higher recall on core-level faults but
cannot detect link faults and misclassify propagated slowdowns as root
causes.
This produces volatile results: Perseus achieves 6.56\% FPR on
VGG but 91.71\% on ResNet-50, where deeper residual paths
create longer propagation chains.

\project achieves 69--87\% accuracy with 5--26\% FPR across all
workloads, and no single baseline exceeds 70\% accuracy on more than
two workloads.
This consistency stems from two design properties: SL-Tracer's
multi-level communication graph jointly models core and link faults
(addressing scope), and FailRank's iterative scoring separates root
causes from backpressure victims (addressing topology).
Importantly, \project targets diagnostic localization:
an average 13.23\% FPR substantially narrows the candidate set from many cores to a ranked shortlist, which is the operationally significant improvement over having no on-chip detection capability.

Workload structure affects difficulty but \project remains robust.
Binary Tree yields the highest accuracy (86.69\%) due to its regular
communication patterns.
DarkNet-19 is the hardest case in the traditional networks: its uniform workload mapping produces
correlated traces that compress the performance gap between healthy
and degraded components, limiting even \project to 74.59\%.
Workloads with diverse layer structures (VGG)
consistently yield 82\% accuracy.
On Early-Exit ResNet, SLOTH achieves a detection accuracy of 81.24\%, outperforming the standard ResNet-50 (74.86\%). This improvement is driven by a significant drop in FPR. The early-exit branches introduce shorter and simpler execution flows, significantly easing the detection.
As for Qwen2.5-0.5B, we evaluate it on an $8 \times 8$ mesh topology rather than $4 \times 4$, since the latter lacked sufficient capacity to map it. 
Since the frequent global synchronizations exacerbate the fail-slow propagation and memory-bound dataflow reduces available candidate computation/communication patterns,
Qwen2.5 yields the lowest accuracy (69.68\%) and highest FPR (26.24\%) among all tested workloads for \project. However, \project still outperforms all baseline algorithms.

Perseus performs a little better on VGG. The main reason is that its architecture consists almost entirely of dense FC and large Conv layers with no branching, skip connections, or complex data redistribution, producing minimal NoC traffic variation. Perseus focuses on compute-core anomalies only and therefore is well-matched to this workload. However, SLOTH's topology-aware detection activates precisely when communication patterns carry diagnostic signal. On VGG, there is little such signal to exploit.

\begin{figure}[tbp]
  \centering
  \includegraphics[width=\linewidth]{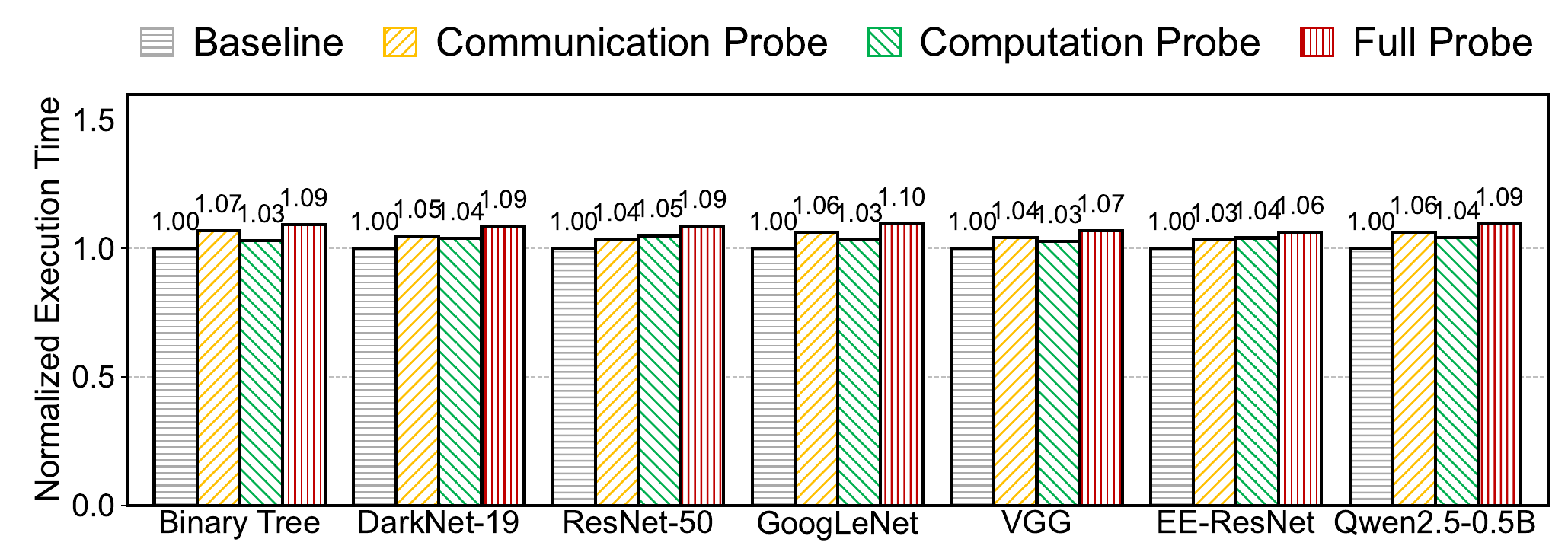}
  \caption{Time overhead of SL-Compiler probing.}
  \label{fig:time-overhead}
  \vspace{-10pt}
\end{figure}

\subsection{Overhead Analysis}
\label{sec:overhead}

\noindent\textbf{Time Overhead.}
\label{sec:time-overhead}
We evaluate the runtime overhead of SL-Compiler's probe-based data
collection under three configurations: Communication Probe
(communication instructions only), Computation Probe (computation
instructions only), and Full Probe (all instructions), using
probe-free execution as the baseline.

Figure~\ref{fig:time-overhead} shows that even under the most
extensive instrumentation (Full Probe), execution time increases by
at most 10\% across all workloads.
For targeted scenarios such as Communication Probe, overhead remains
below 7\%.
This confirms the less than 10\% overhead claim from SL-Compiler's
contribution.

\noindent\textbf{Storage Overhead.}
\label{sec:mem-overhead}
We evaluate SL-Recorder's storage efficiency by comparing memory consumption before and after compression. The raw format records full instruction details (index, timestamps, operands, ...) and serves as the baseline. For each workload, we separately measure communication and computation trace sizes, then apply compression to quantify storage savings. We also compare against IASO (communication traces), Perseus and ADR (computation traces)' memory requirement to demonstrate SL-Recorder's advantage.

\begin{figure}[tbp]
  \centering
  \includegraphics[width=\linewidth]{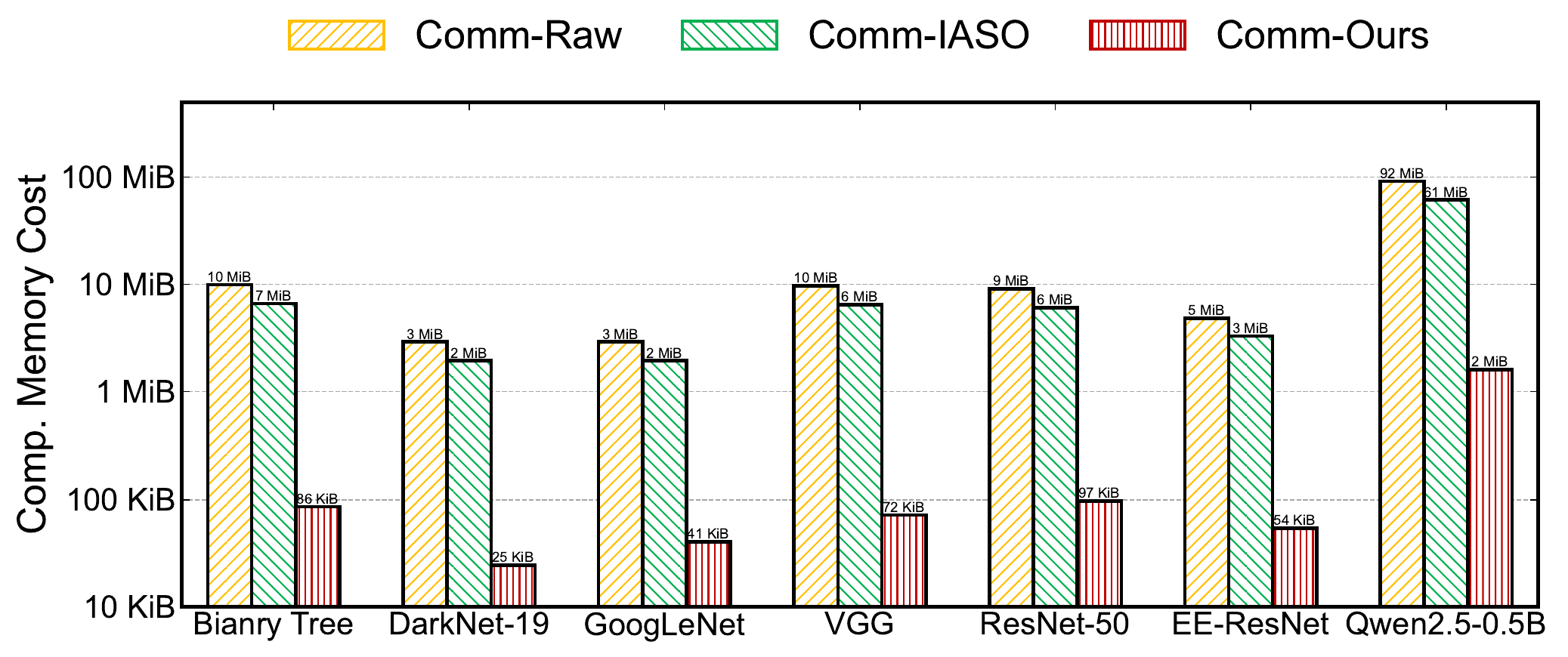}
  \captionsetup{skip=2pt}
  \caption{Memory cost of communication traces with no optimization, IASO and SL-Recorder.}
  \label{fig:comm-memory-save}
\end{figure}

\begin{figure}[htbp]
  \centering
  \includegraphics[width=\linewidth]{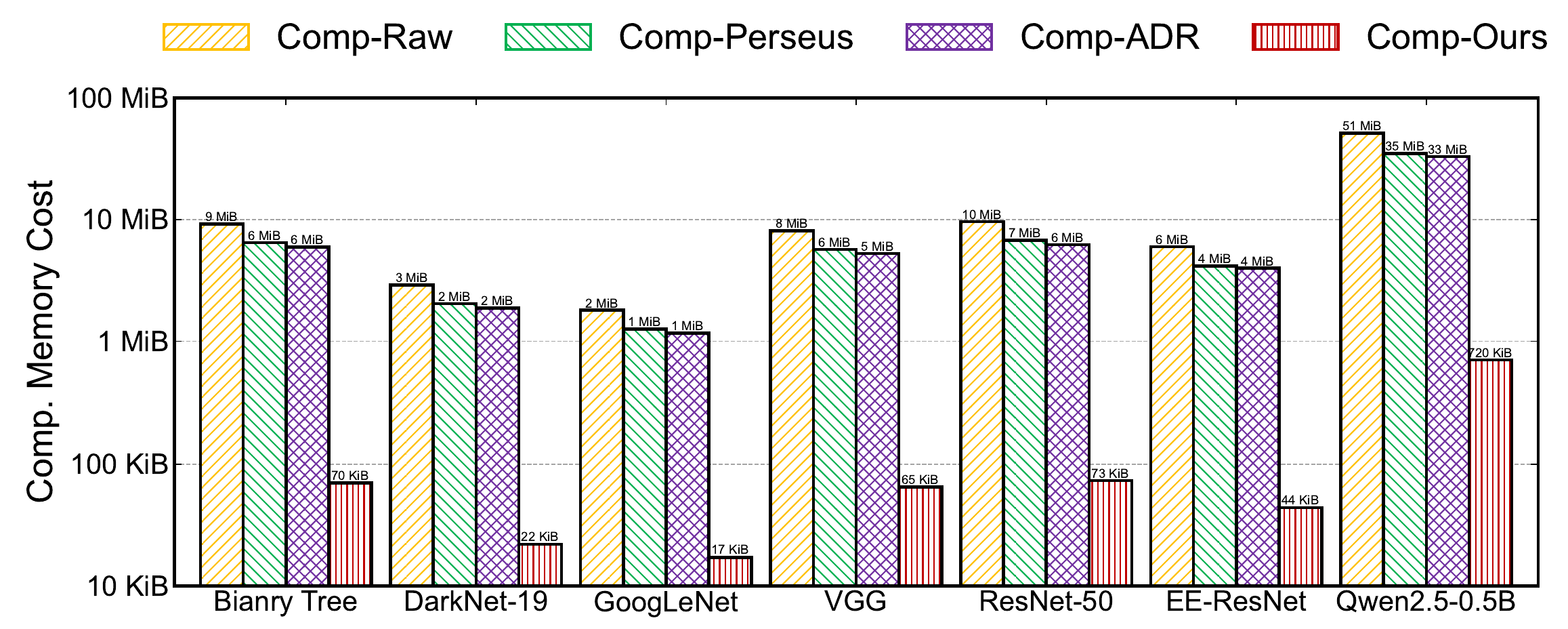}
  \captionsetup{skip=2pt}
  \caption{Memory cost of computation traces with no optimization, Perseus, ADR and SL-Recorder.}
  \label{fig:comp-memory-save}
  % \vspace{-15pt}
\end{figure}

Figure~\ref{fig:comm-memory-save} shows that SL-Recorder compresses
communication traces by over 100$\times$ across most workloads (e.g.,
Binary Tree: \SI{10}{\mebi\byte} raw $\to$ \SI{86}{\kibi\byte}).
IASO provides only 30--40\% reduction because it retains full
communication traces.
Figure~\ref{fig:comp-memory-save} shows similar results for
computation traces: SL-Recorder achieves 70--99\% compression, while
Perseus and ADR achieve only 25--50\% as they retain per-instruction
records.
Overall, SL-Recorder reduces trace storage by over two orders of
magnitude, bringing per-core storage within the kilobyte-scale SRAM
budget identified in the introduction.

\begin{figure}[tbp]
  \small
  \centering
  \begin{subfigure}[b]{0.32\linewidth}
    \includegraphics[width=\linewidth]{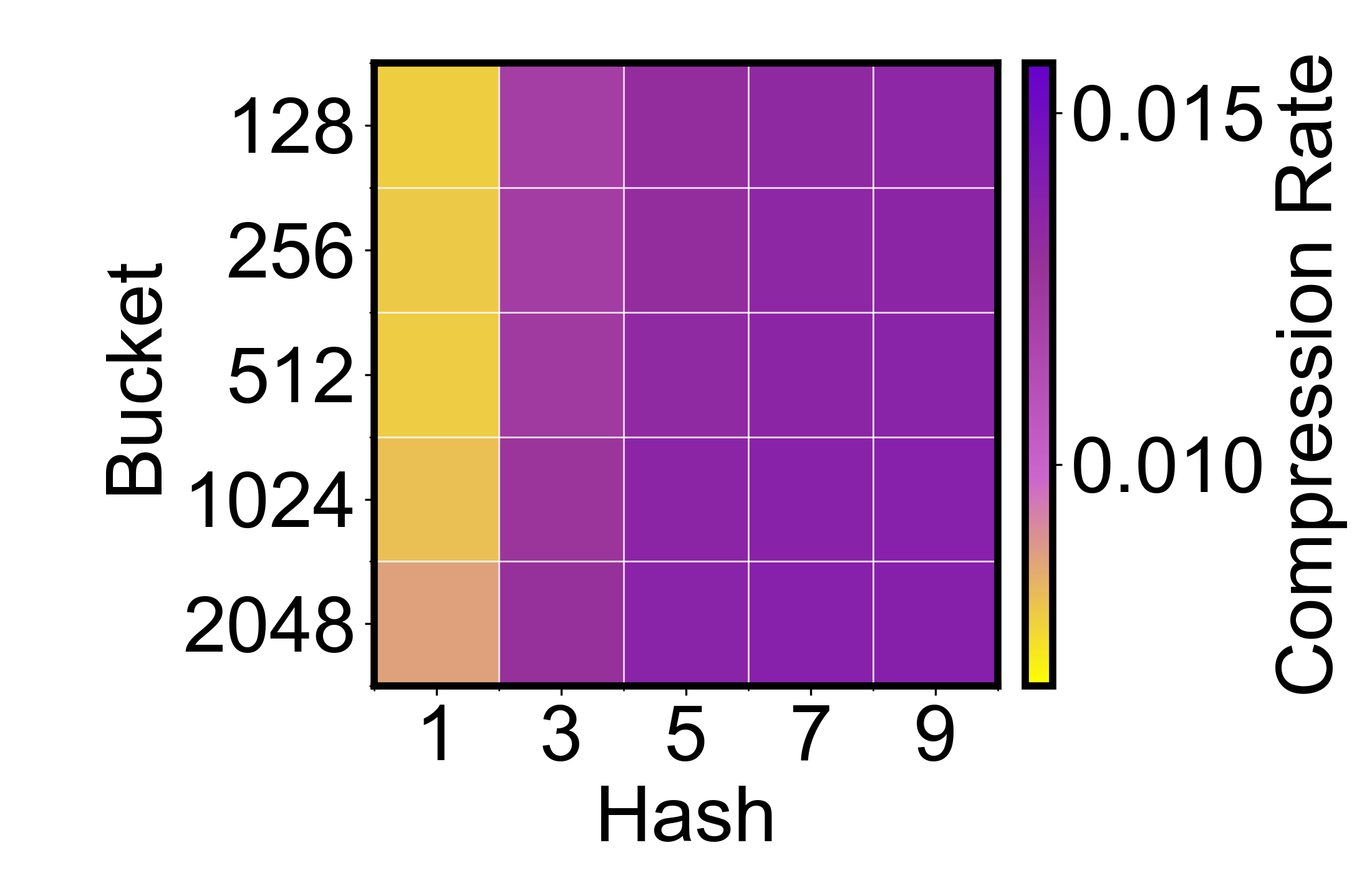}
    \captionsetup{skip=0pt}
    \caption{S=8192,T=10}
    \label{fig:compression-a}
  \end{subfigure}
  \begin{subfigure}[b]{0.32\linewidth}
    \includegraphics[width=\linewidth]{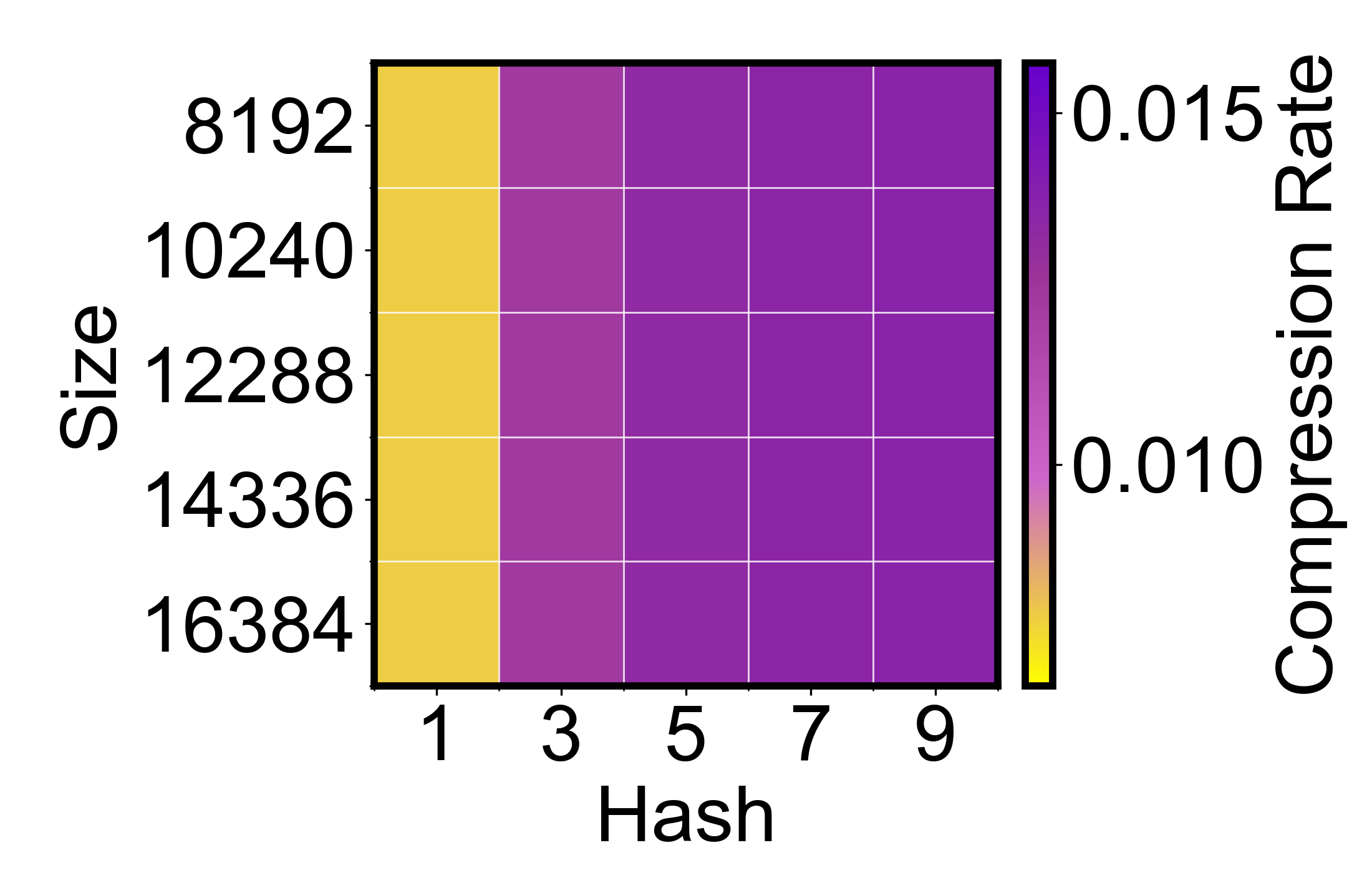}
    \captionsetup{skip=0pt}
    \caption{B=512,T=10}
    \label{fig:compression-b}
  \end{subfigure}
  \begin{subfigure}[b]{0.32\linewidth}
    \includegraphics[width=\linewidth]{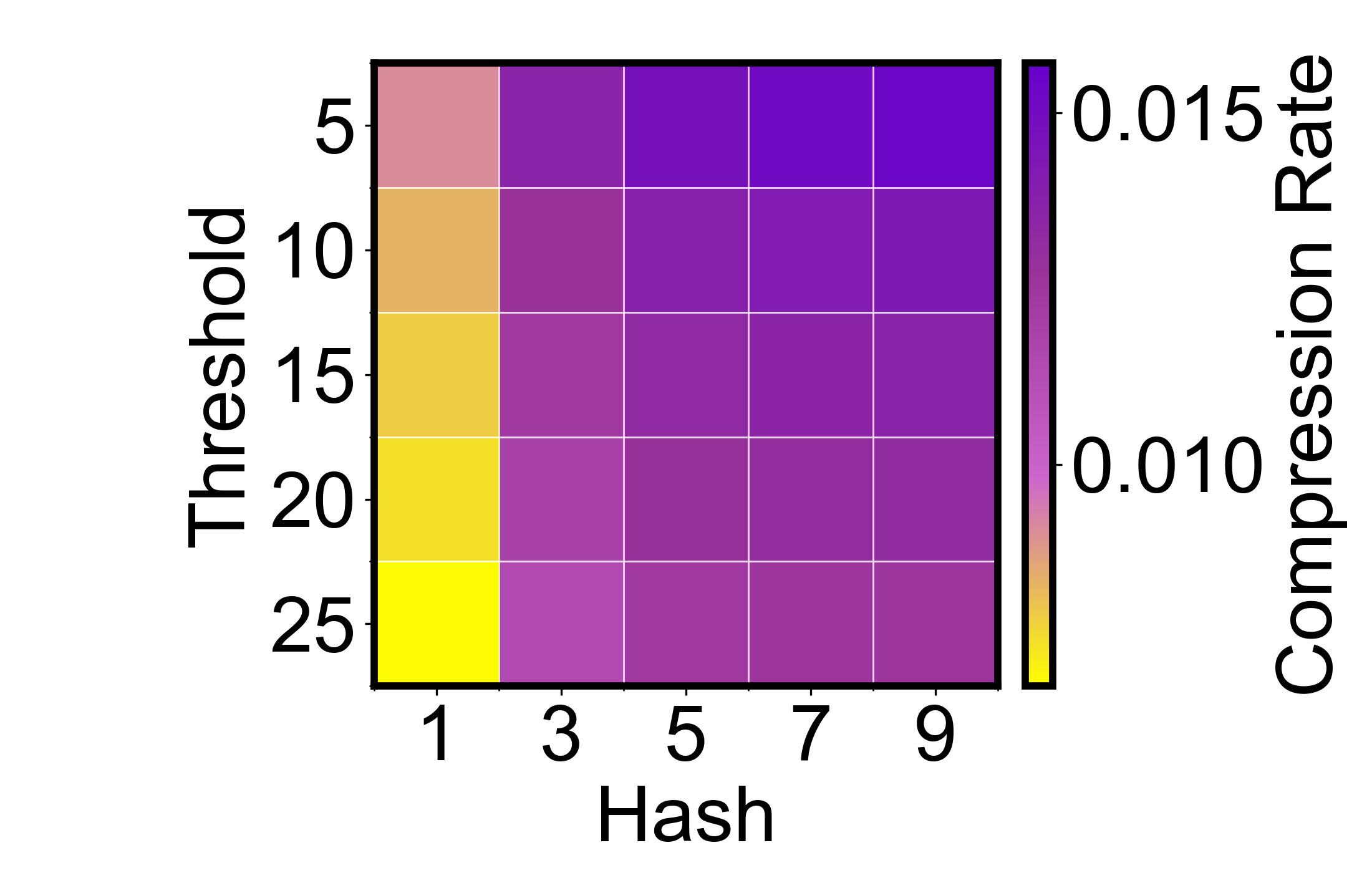}
    \captionsetup{skip=0pt}
    \caption{B=512,S=8192}
    \label{fig:compression-c}
  \end{subfigure}

  \vspace{8pt}

  \begin{subfigure}[b]{0.32\linewidth}
    \includegraphics[width=\linewidth]{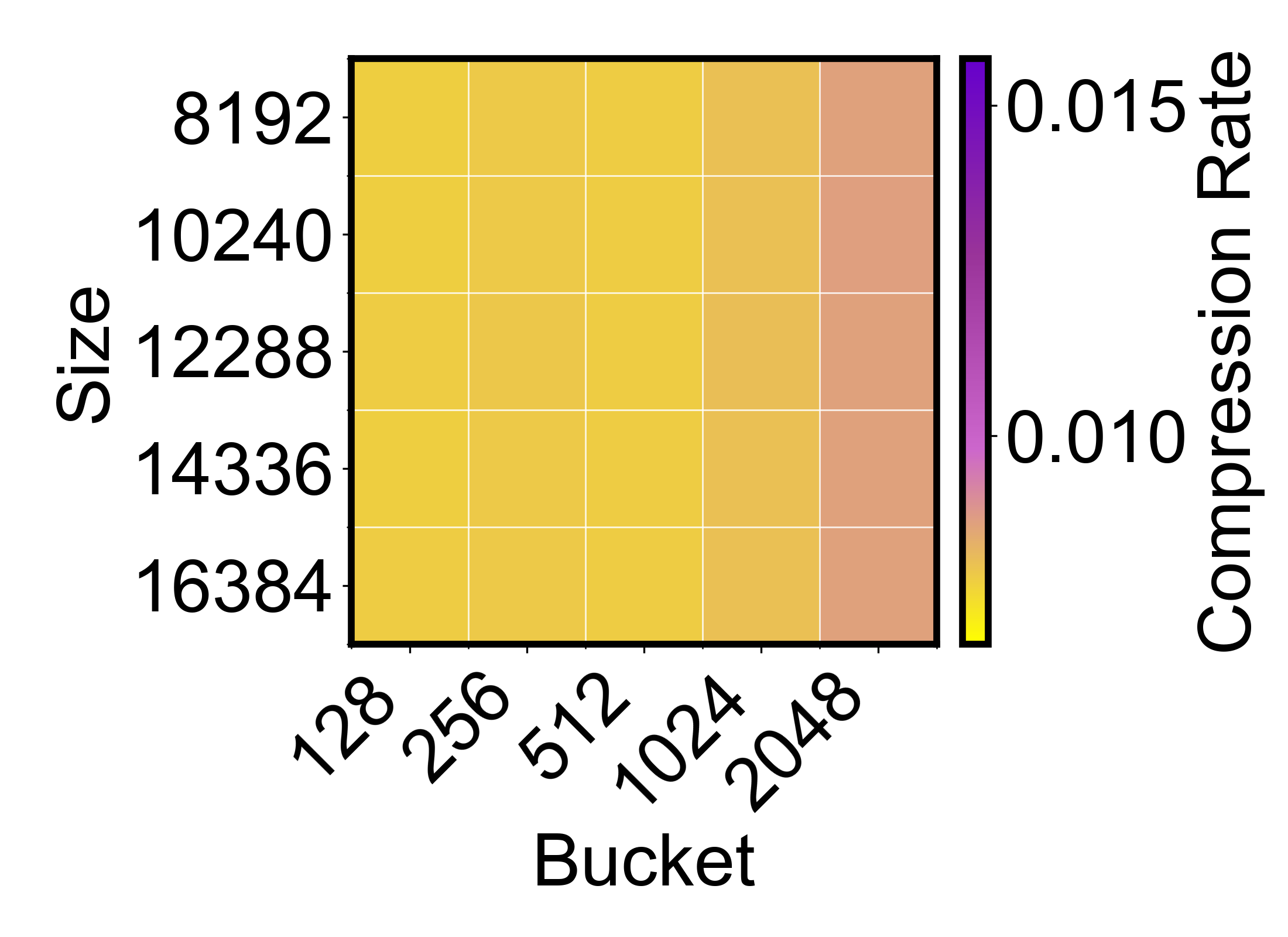}
    \captionsetup{skip=0pt}
    \caption{H=3,T=10}
    \label{fig:compression-d}
  \end{subfigure}
  \begin{subfigure}[b]{0.32\linewidth}
    \includegraphics[width=\linewidth]{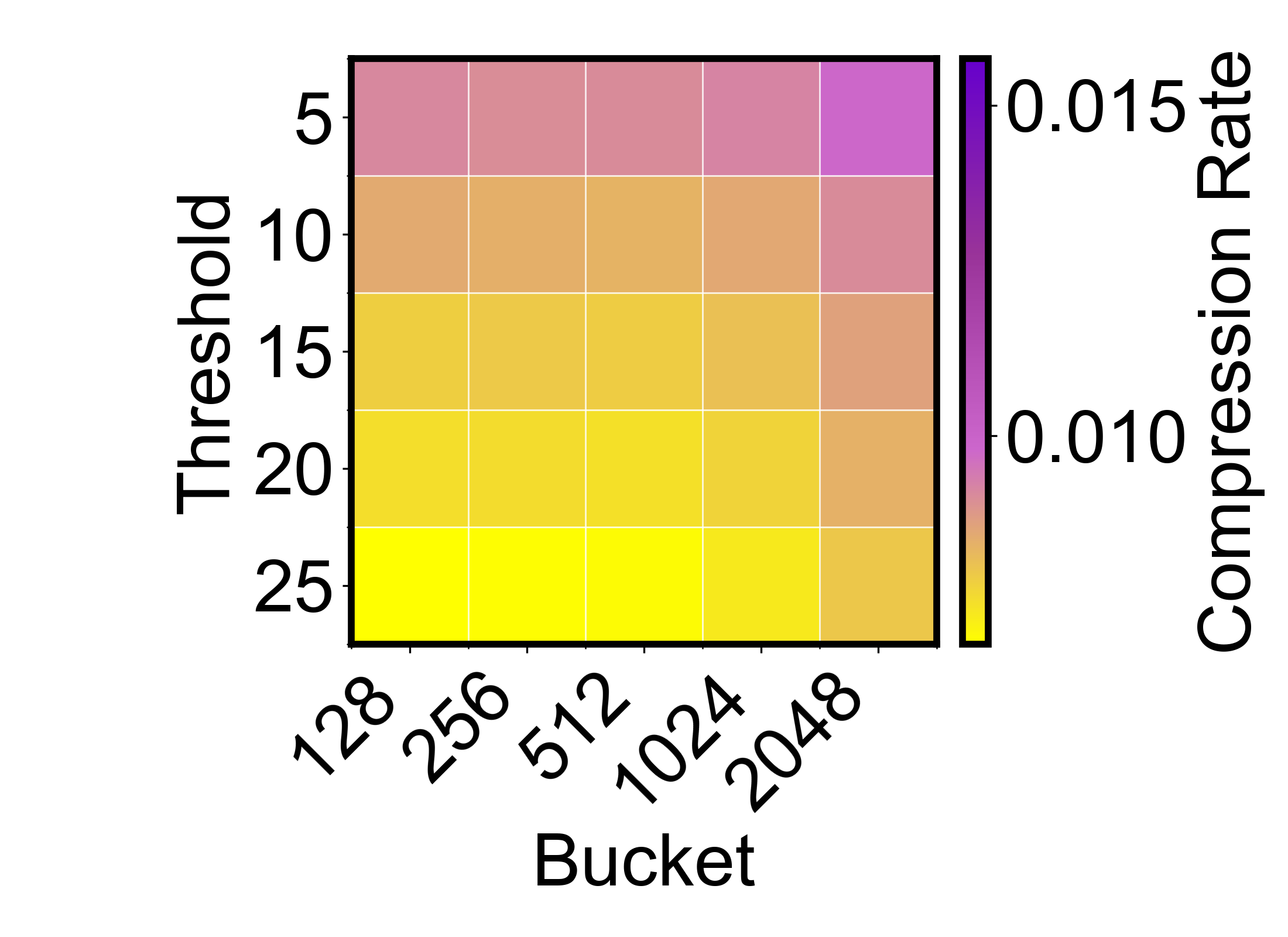}
    \captionsetup{skip=0pt}
    \caption{H=3,S=8192}
    \label{fig:compression-e}
  \end{subfigure}
  \begin{subfigure}[b]{0.32\linewidth}
    \includegraphics[width=\linewidth]{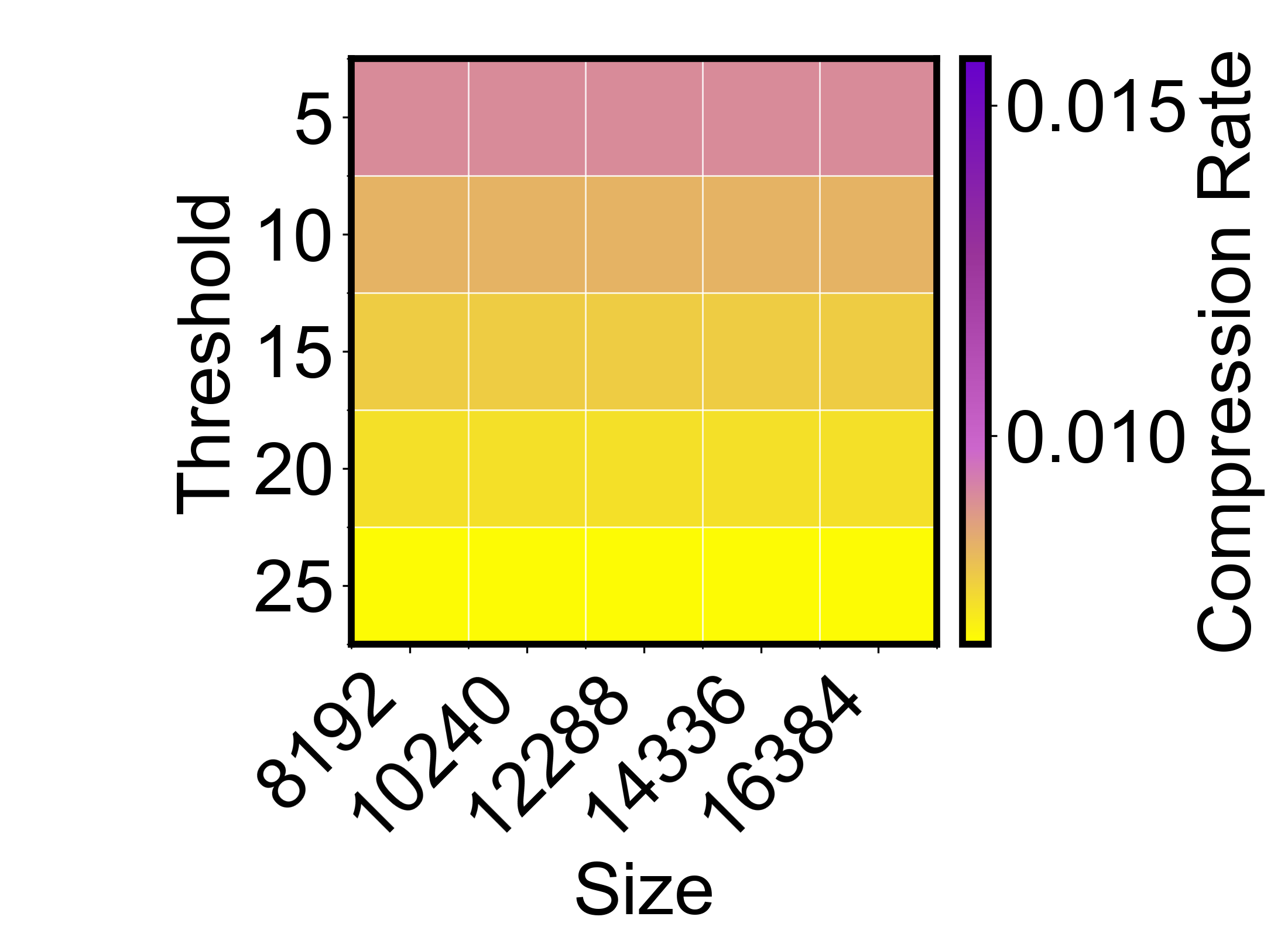}
    \captionsetup{skip=0pt}
    \caption{H=3,B=512}
    \label{fig:compression-f}
  \end{subfigure}
  \vspace{9pt}
  \captionsetup{skip=1pt}
  \caption{SL-Recorder compression rate: Impact of the Hash (H), Bucket (B), Size (S), and Threshold (T).}
  \label{fig:compression-rate}
  \vspace{-10pt}
\end{figure}

We further examine how SL-Recorder's four internal parameters
(\texttt{Hash H}, \texttt{Bucket B}, \texttt{Size S},
\texttt{Threshold T}) affect compression.
Figure~\ref{fig:compression-rate} shows pairwise heatmaps on
DarkNet-19, fixing two parameters at defaults and varying the other
two.

Figure \ref{fig:compression-rate} illustrates the impact of different parameter settings on the compression ratio of SL-Recorder. Among the four parameters, \texttt{H}, \texttt{B}, and \texttt{T} show a clear influence on compression effectiveness, while \texttt{S} has little impact. Specifically, a smaller number of hash functions and buckets, as well as a larger threshold, consistently lead to stronger compression. In contrast, varying \texttt{S} across a wide range does not significantly alter the compression ratio, indicating that this parameter is relatively insensitive. These observations suggest that tuning hash, bucket, and threshold values is crucial for achieving high compression efficiency, whereas Stage-2 size can be fixed without noticeably affecting performance.

\noindent\textbf{Accuracy--Overhead Tradeoff.}
\label{sec:dse}
To find the optimal SL-Recorder configuration, we conduct a design
space exploration across four parameters: hash functions
(\texttt{Hash}), buckets (\texttt{Bucket}), Stage-2 size
(\texttt{Size}), and pattern-selection threshold (\texttt{Threshold}).
These jointly determine compression efficiency and detection accuracy.

We define a composite optimization objective as:
$COST = ACC^{\alpha} \times R^{\beta} \times M^{\gamma}$,
where $ACC$ represents the detection accuracy, $R$ denotes the data compression rate, and $M$ refers to the memory overhead of data structure. By adjusting the hyperparameters $\alpha,\beta,\gamma$, users can flexibly prioritize accuracy, compression rate, or memory overhead according to their own deployment requirements. Here we set $\alpha=-1,\beta=\gamma=1$, considering all three objectives equally.

\begin{figure}[htbp]
  \centering
  \begin{subfigure}[b]{0.47\linewidth}
    \centering
    \includegraphics[width=\linewidth]{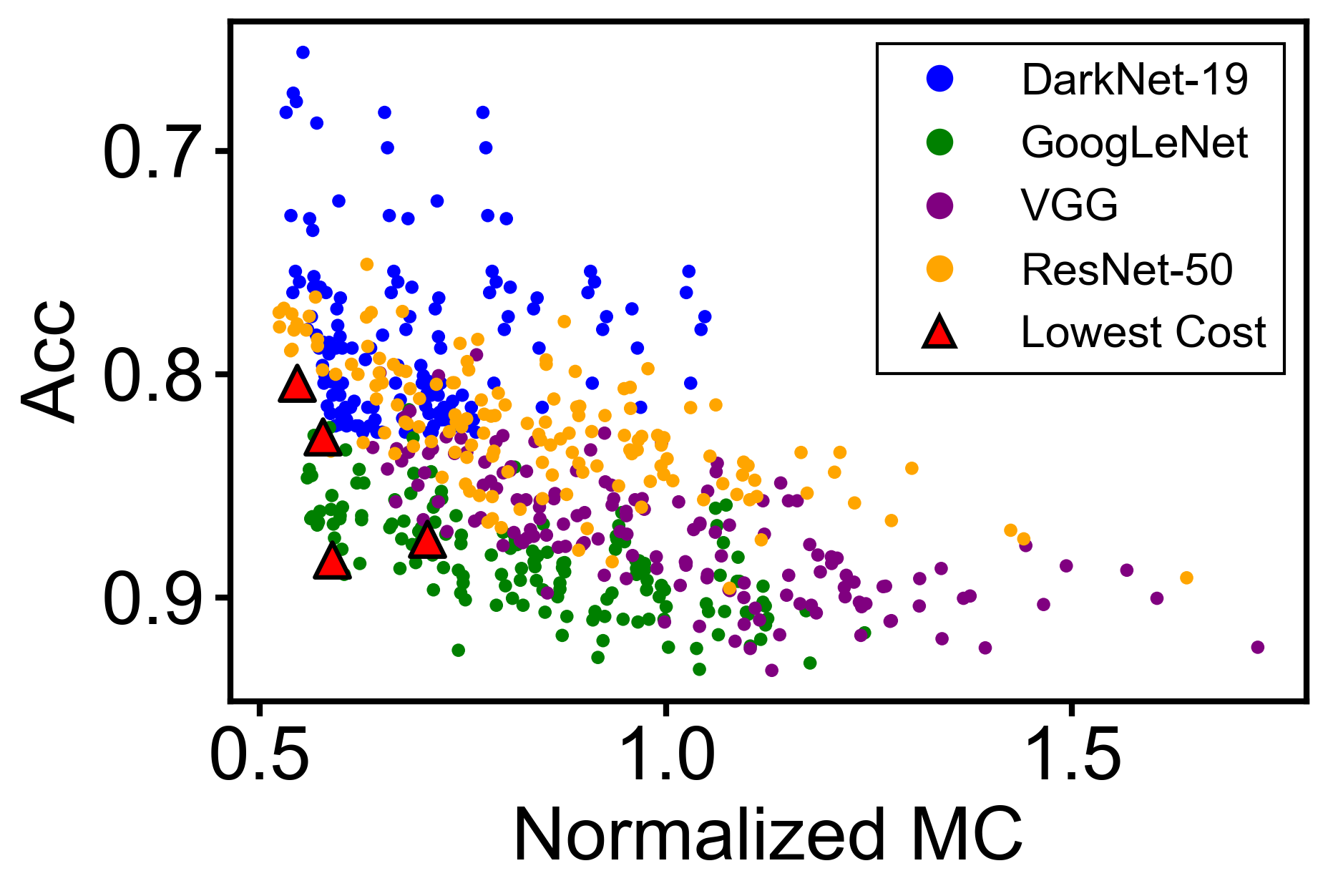}
    \captionsetup{skip=0pt}
    \caption{4$\times$4 Mesh.}
    \label{fig:dse}
  \end{subfigure}
  \hfill
  \begin{subfigure}[b]{0.47\linewidth}
    \centering
    \includegraphics[width=\linewidth]{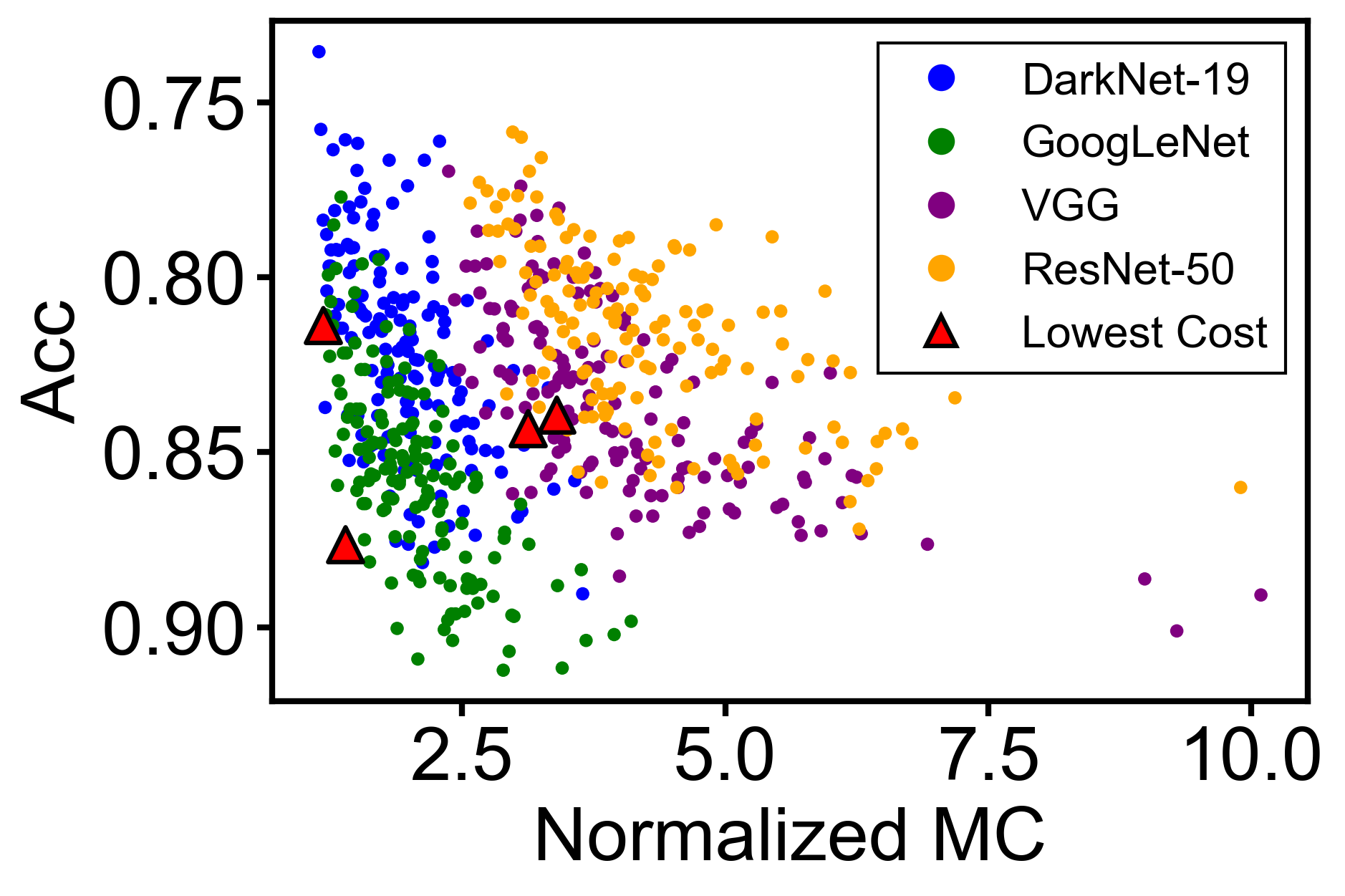}
    \captionsetup{skip=0pt}
    \caption{8$\times$8 Mesh.}
    \label{fig:dse8-8}
  \end{subfigure}
  \vspace{9pt}
  \captionsetup{skip=-1pt}
  \caption{SL-Recorder Design Space Exploration.}
  \label{fig:DSE}
  \vspace{-12pt}
\end{figure}

Figure~\ref{fig:dse} shows the results on a 4$\times$4 accelerator,
where each point is a parameter configuration and colors represent
workloads.
The design space exhibits clear trade-offs: lower memory cost often
degrades accuracy, while moderate-cost configurations yield balanced
performance.
Pareto-optimal cases are highlighted with red triangles.
Figure~\ref{fig:dse8-8} shows that the 8$\times$8 accelerator
exhibits the same trade-off structure, indicating that optimal
parameter choices transfer across scales.

\begin{figure}[tbp]
  \centering
  \includegraphics[width=\linewidth]{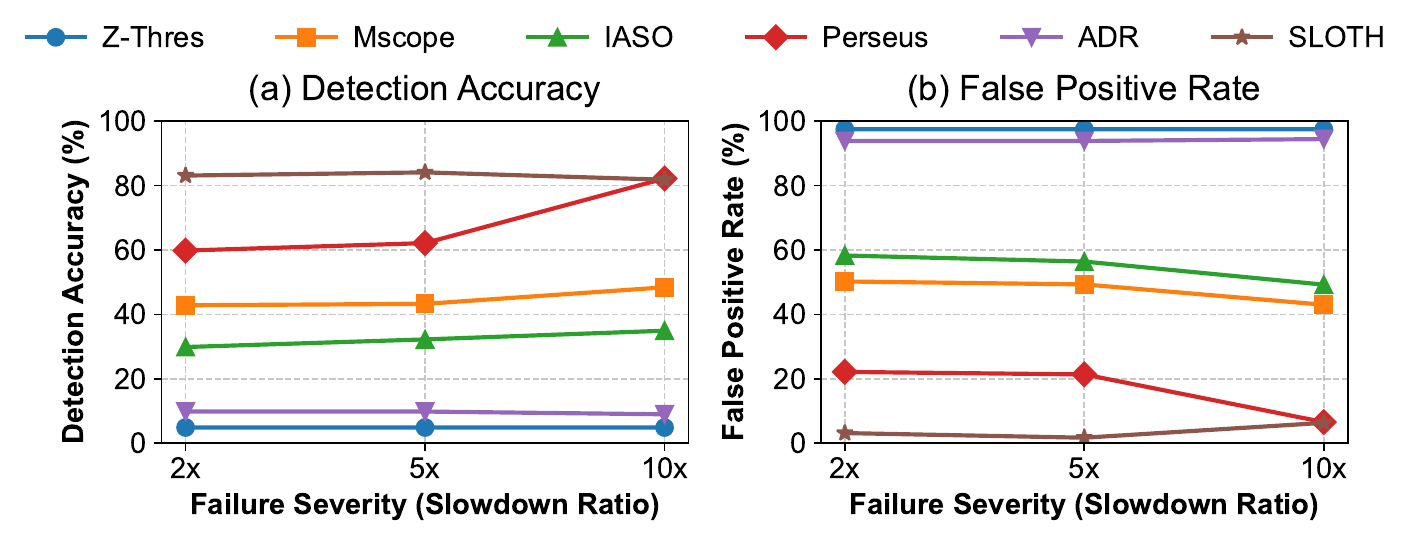}
  \captionsetup{skip=2pt}
  \caption{Evaluation of fail-slow localization on VGG under varying degradation severities. (a) Detection accuracy and (b) false positive rate under 2$\times$, 5$\times$, and 10$\times$ slowdown ratios.}
  \label{fig:sensitivity}
  \vspace{-10pt}
\end{figure}

\subsection{Sensitivity Analysis}
\label{sec:sensitivity}

To evaluate the robustness of \project under different failure severities comparing with other baselines, we conduct a sensitivity analysis by varying the injected slowdown ratio. Specifically, we configure the fail-slow setting to enforce 2$\times$, 5$\times$, and 10$\times$ latency degradations on the affected cores or links. These three severity configurations are tested on VGG.

As shown in Figure~\ref{fig:sensitivity}, for most baselines, diagnosing milder fail-slow failures is more challenging. As the failure severity decreases from 10$\times$ down to 2$\times$, the overall detection accuracy exhibits a slight degradation while the FPR generally increases.
Despite this inherent difficulty, \project demonstrates strong robustness. Even under the highly challenging 2$\times$ slowdown scenario, the framework successfully maintains an accuracy of over 81\% and keeps the FPR below 7\% across all evaluated workloads.

Furthermore, we observe a distinct behavioral pattern in Perseus compared to other baselines. While most methods show marginal performance changes across different slowdown ratios, Perseus's effectiveness is highly dependent on the failure severity. It performs well at the 10$\times$ severity but suffers a precipitous drop in accuracy when the slowdown ratio decreases to 5$\times$ and 2$\times$ (from 82.20\% to 59.82\%). The reason of this vulnerability probably lies in Perseus's reliance on linear regression for failure detection, milder 2$\times$ fail-slow failures make linear assumptions fail to encapsulate the natural variations of the accelerator. Consequently, the linear regression model fails to establish a clear decision boundary to distinguish subtle hardware fail-slow faults from benign system fluctuations, leading to a sharp decline in overall robustness.

\begin{figure}[tb]
  \centering
  \includegraphics[width=\linewidth]{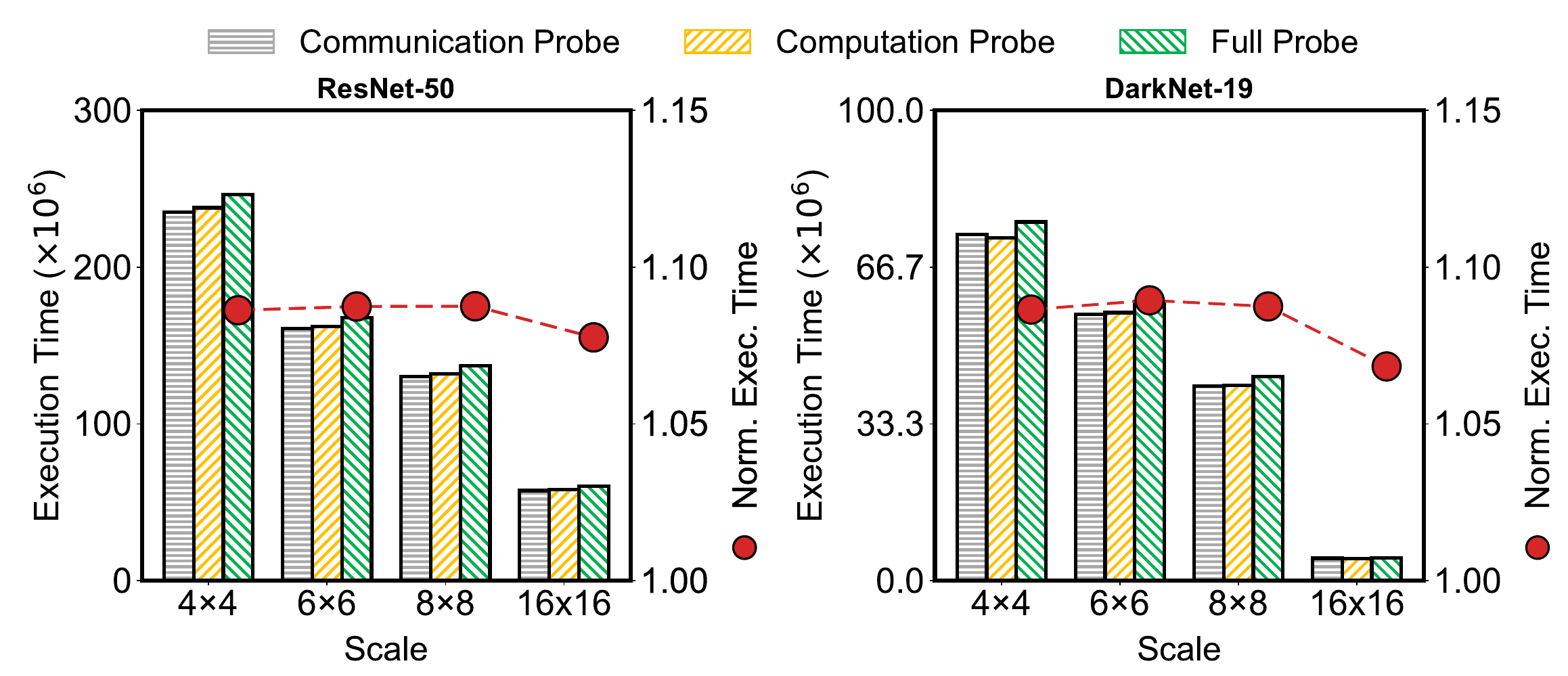}
  \captionsetup{skip=-1pt}
  \caption{Scalability of \project: Time overhead.}
  \label{fig:scalability-time}
  \vspace{-10pt}
\end{figure}

\begin{figure}[t]
  \centering
  \includegraphics[width=\linewidth]{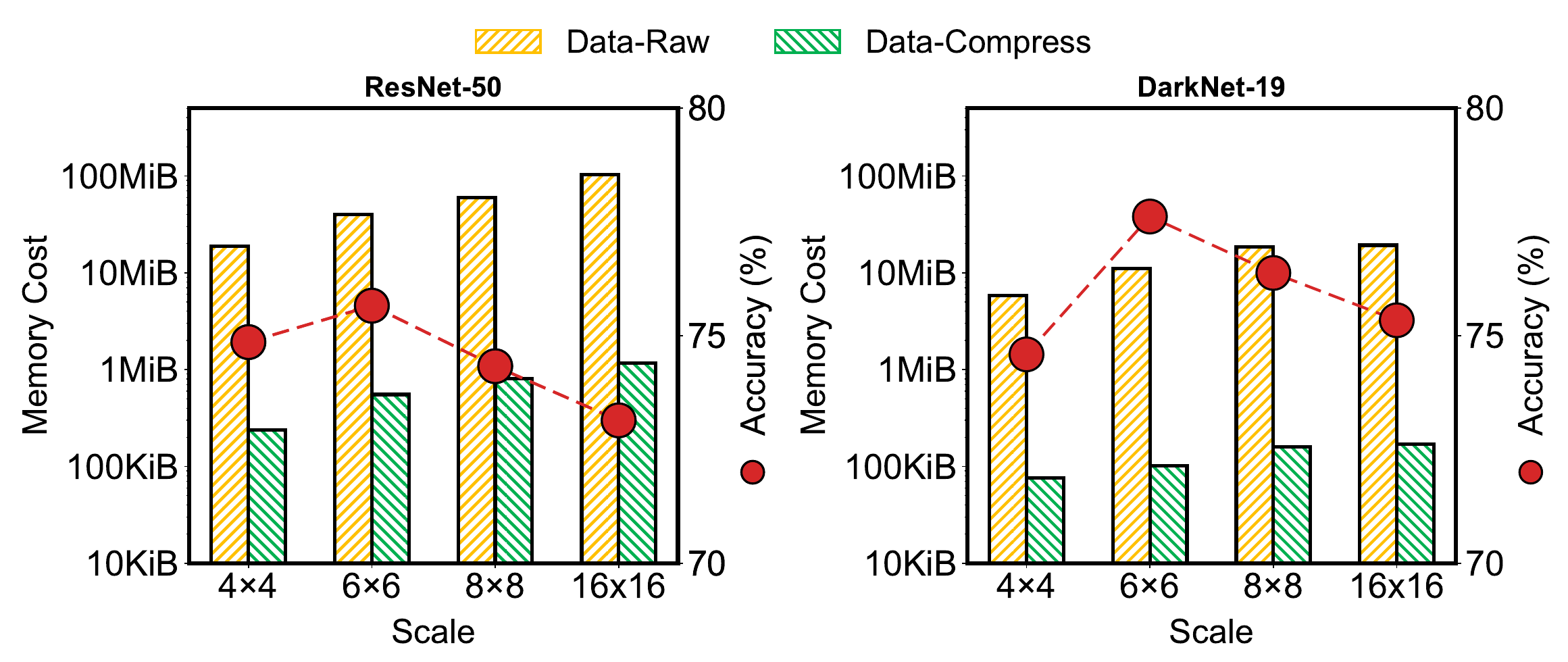}
  \captionsetup{skip=-1pt}
  \caption{Scalability of \project: Memory cost and detection accuracy.}
  \label{fig:scalability-memory}
  \vspace{-10pt}
\end{figure}

\subsection{Scalability Analysis}
\label{sec:scalability}

We test \project on 4$\times$4, 6$\times$6, 8$\times$8, and 16$\times$16
accelerators running ResNet-50 and DarkNet-19, measuring execution
time under three probing modes and tracking memory cost and detection
accuracy across scales.

Figure~\ref{fig:scalability-time} shows that probing overhead remains
consistently low as scale grows: the normalized overhead of the full
probe stays below 10\% at all sizes.
Figure~\ref{fig:scalability-memory} shows that while raw trace volume
grows with scale, SL-Recorder maintains nearly 100$\times$
compression across all configurations.
Detection accuracy remains stable with no degradation as the
architecture scales, confirming that \project's pipeline does not
degrade with increasing core count.

To confirm that \project is not specific to mesh networks, we
additionally evaluate DarkNet-19 on ring, torus, and dragonfly
topologies.
We choose DarkNet-19 because it is the most challenging workload
(\S\ref{sec:detection}).
Detection accuracy and FPR remain consistent across all four
topologies (69--75\% accuracy, 19--27\% FPR),
confirming that \project's detection pipeline depends on the
computation graph and communication dependencies, not the specific
interconnect structure.

\section{Discussion}

As the first on-chip fail-slow detection framework for spatial DNN accelerators, \project bridges the gap between trace monitoring and fine-grained root cause analysis. In this section, we outline the practical implications, the extensibility to other platform of \project, and possible future research directions.

\noindent\textbf{Fail-Slow Mitigation.} A key motivation for pushing fail-slow detection to the intra-chip granularity is to enable further mitigation. Traditional node-level or chip-level detection merely flags a chip for replacement, which is highly inefficient for modern large-scale accelerators. By pinpointing exactly which core or NoC link is experiencing performance degradation, \project enables practical mitigation strategies. At the compiler level, the runtime scheduler can dynamically regenerate task schedules and data layouts. By marking degraded cores as unfavorable, the mapping engine can route critical-path operators away from them, effectively hiding the fail-slow failure in subsequent inference batches. At the microarchitectural level, if the root cause is isolated to a link, the NoC controller can employ adaptive routing mechanisms to bypass the affected path, redistributing traffic across healthy interconnects without taking the entire chip offline.

\noindent\textbf{Integration with Other Platforms.} While \project is designed for spatial DNN accelerators, its modular architecture ensures broad extensibility to other spatial architectures, such as CGRAs or distributed systems. Porting \project requires adapting three platform-specific interfaces. First, SL-Compiler relies on the dataflow IR. Adapting the compiler simply requires aligning its parsing logic with the target platform's specific IR format. Second, the online sketching mechanism in SL-Recorder operates on abstract trace entries with a fixed schema. Integrating it with other platforms only requires matching the target's underlying trace sequences to this standardized input footprint. Third, for root-cause analysis, SL-Tracer exploits the hardware topology and the logic dependencies. It's easy to work on other interconnects by updating the physical hardware topology definition during the MCG construction phase, without changing the core inference algorithm.

\noindent\textbf{Future Work.} While \project resolves the lack of fine-grained visibility in spatial accelerators, scaling it to massive, multi-chiplet, or wafer-scale systems remains a problem, and will be a direction for future work. As systems scale to tens of thousands or even millions of cores, new hardware behaviors may emerge, such as hierarchical NoC topologies, lossy die-to-die interconnects, and significant non-uniformity in communication delays that complicate the detection. Future efforts should adapt the detection algorithm to handle hierarchical, multi-die propagation boundaries and coordinate distributed instances across chiplets while strictly preserving memory efficiency. Furthermore, as models increasingly rely on highly unpredictable data-dependent control flows and dynamic operator offloading, evolving instrumentation to track sparse dataflow and runtime task migration represents an important next step in on-chip failure diagnosis.

% \vspace{10pt}
\section{Related Work}

\textbf{Distributed Tracing}. Existing distributed tracing systems reduce overhead through reactive collection or selective storage\cite{10.1145/3132747.3132749,10.1145/3357223.3362736,10.1145/3267809.3267841,36356}, but these approaches are often not suitable for diagnosing fail-slow behaviors. For instance, Pivot Tracing \cite{10.1145/2815400.2815415} enables on-demand instrumentation based on user queries, yet critical early-stage evidence of gradual degradation is often missed. Hindsight \cite{285115} uses post-hoc triggers to commit buffered traces, but the performance decay of fail-slow issues may not activate these triggers in time, leading to the loss of crucial data. Similarly, while Sieve \cite{9590295} applies an attention model to prioritize traces, it is biased towards conspicuous spikes and can overlook the weak, temporally-correlated signals characteristic of a fail-slow. In contrast, SL-Compiler takes a proactive, structure-aware approach:
it analyzes the computation graph at compile time to identify
operator boundaries suited for fail-slow diagnosis,
capturing degradation evidence continuously
rather than reacting to triggers after the fact.

\noindent\textbf{Trace Storage Optimization.} The sheer volume of trace data needs optimizations for efficient storage\cite{10.1109/ASE.2019.00085,10.1145/3669940.3707287,273737}. For example, Mint \cite{10.1145/3669940.3707287} reduces storage footprint by decomposing traces into common patterns and specific attributes. However, this structural compression is agnostic to performance semantics; it does not help in identifying or prioritizing traces that exhibit gradual latency degradation. Other works leverage sketches for data summarization\cite{10.1145/3589334.3645581,9064148,10.1109/TKDE.2022.3223686}. For example, Dynamic Sketch \cite{9064148} uses a doorkeeper mechanism to efficiently detect heavy hitters, but it focus on request frequency and is misaligned with fail-slow analysis. Similarly, while SketchLib \cite{276934} provides highly optimized sketch implementations for Reconfigurable Match-Action Tables (RMT), its design is incompatible with the stringent on-chip memory constraints. Our SL-Recorder addresses a fundamentally different detection target:
whereas existing methods summarize frequent patterns,
SL-Recorder's two-stage compression design
tracks per-bucket latencies
to isolate the infrequent deviations that signal fail-slow degradation,
while fitting within on-chip SRAM constraints.

\noindent\textbf{Failure Detection and Root Cause Analysis.} Ensuring system reliability requires robust mechanisms for both failure detection and subsequent root cause analysis (RCA)\cite{groot,sleuth,sage,jin2020scalana}. For instance, ScalAna \cite{jin2020scalana} leverages backtracking on a Program Structure Graph (PSG) to localize root causes. However, its reliance on a purely logical graph, which abstracts away the hardware topology, hinders the diagnosis of communication-related bottlenecks. Another class of methods focuses on component-level metrics\cite{panda2019iaso,10.5555/3585938.3585942,10.5555/3767955.3767975}. Approaches like IASO\cite{panda2019iaso}, using timeout signals and peer-scoring, and ADR \cite{10.5555/3767955.3767975}, which monitors latency and frequency shifts, are effective at identifying node-level slowdowns. Nevertheless, their node-centric perspective fails to capture performance degradation occurring on the communication links themselves. Our SL-Tracer overcomes these limitations using a Multi-level Communication Graph (MCG) that integrates software communication dependencies with the underlying physical topology. By applying an iterative FailRank algorithm on MCG, SL-Tracer can pinpoint the root cause of fail-slow, irrespective of whether the origin is a computational node or a communication link.

\vspace{5pt}
\section{Conclusion}

In this work, we introduce \project, a tracing and analyzing framework for detecting and localizing fail-slow failures in spatial DNN accelerators. \project provides a full-stack solution that spans lightweight data collection, online trace compression, and graph-based root cause analysis. Specifically, we design SL-Compiler to insert low-overhead tracing probes into DNN workload, SL-Recorder to achieve runtime trace filtering and compression, and SL-Tracer for accurate fail-slow localization. We evaluate \project using a set of representative networks and hardware configurations. Experiments demonstrate the effectiveness of \project. It reduces the storage overhead of detection traces by an average of 115.9$\times$, while achieving an average fail-slow detection accuracy from 69.68\% to 86.69\%.
As the first framework for on-chip fail-slow detection, \project
opens directions for future work, including characterizing how hardware parameters, workload communication patterns, and emerging operator topologies jointly shape fail-slow propagation and detection accuracy.

\appendix
\section{Artifact Appendix}

\subsection{Abstract}

This appendix describes the setup and instructions to reproduce the experimental results in the paper. The artifact provides the materials used to evaluate SLOTH on simulated
spatial DNN accelerators. It includes the simulator and analysis code,
architecture descriptions, workload mappings, fail-slow injection cases,
baseline implementations, experiment scripts, and plotting scripts. The
packaged experiments cover the paper's accuracy and false-positive-rate
comparison, trace-size and execution-overhead measurements, fail-slow severity
sensitivity study, and accelerator scale study.

\subsection{Artifact check-list (meta-information)}

{
\begin{itemize}
  \item {\bf Algorithm:} SL-Compiler, SL-Recorder,
  multi-level communication graph construction, FailRank, and baselines
  Z-Thres, Mscope, IASO, Perseus, and ADR.
  \item {\bf Model:} Binary Tree, DarkNet-19, GoogLeNet, VGG, ResNet-50,
  Early-Exit ResNet, and Qwen2.5-0.5B.
  \item {\bf Data set:} Small example JSON inputs,
  architecture descriptions in \verb|data/archs|,
  mapping JSON files in \verb|data/mappings|, and fail-slow injection
  cases in \verb|data/failslows|.
  \item {\bf Run-time environment:} Python 3.10+. Required Python
  packages are pydantic, simpy, numpy, scipy, scikit-learn, and matplotlib. GNU parallel is optional for parallel
  sweeps.
  \item {\bf Hardware:} Two Intel Xeon Platinum 8375C
  CPUs at 2.90 GHz, 32 cores per socket, two hardware threads per core,
  128 logical CPUs total, and 499 GiB RAM.
  \item {\bf Metrics:} Detection accuracy, FPR, FNR, TP/TN/FP/FN, normalized
  execution-time overhead, raw and compressed trace size, compression rate,
  and recorder memory overhead.
  \item {\bf Output:} Raw and compressed communication
  and computation traces; RCA reports that list predicted
  fail-slow PEs or links, plus aggregate metric and per-case record JSON files.
  \item {\bf Experiments:} Smoke-test simulation and analysis, full trace generation, SLOTH and baseline detection, SL-Recorder design-space exploration, fail-slow
  sensitivity, scale evaluation.
  \item {\bf How much disk space required (approximately)?:} 1TiB.
  \item {\bf How much time is needed to complete experiments (approximately)?:}
  \begin{itemize}
    \item Smoke-test simulation and analysis: within 5 minutes
    \item Full trace generation: 3 days
    \item Root cause detection (Table II): 24 hours
    \item Overhead analysis (Fig. 10, 11, 12): within 5 minutes
    \item Design space exploration (Fig. 13, 14): 2 days per case
    \item Severity analysis (Fig. 15): 10 hours
    \item Scalability analysis (Fig. 16, 17): 5 hours
  \end{itemize}
  \item {\bf Publicly available?:} Yes, on Zenodo.
  \item {\bf Archived (provide DOI)?:} 10.5281/zenodo.21551512
\end{itemize}
}

\subsection{Description}

\subsubsection{How to access}

The artifact is available at 10.5281/zenodo.21551512.

\subsubsection{Hardware dependencies}

\begin{itemize}
    \item The full evaluation requires a powerful CPU platform. More cores and logical CPUs allow more parallel simulation and analysis jobs, which directly reduces evaluation time.
    \item High parallelism also requires large RAM because the simulations and JSON traces can be memory-consuming.
    \item Approximately 1 TiB of storage is required to keep the generated traces, logs, and intermediate results for reproduction.
\end{itemize}

\subsubsection{Software dependencies}

The artifact requires Python 3.10+, pydantic, simpy, numpy, scipy, and scikit-learn. Matplotlib is required for plotting scripts. GNU parallel is optional for large parallel sweeps.

\subsubsection{Data sets}

We provide simulator-ready JSON inputs for all experiments. The
\verb|data/archs| contains accelerator architecture configurations.
The \verb|data/mappings| contains operator-to-core
mappings for the evaluated DNNs. The \verb|data/failslows| contains
fail-slow injection cases, including center, edge-ring, scratch, uniform, and
workload-thermal patterns. The top-level example JSON files provide small
inputs for smoke testing.

\subsubsection{Models}

The evaluated workloads are:
Binary Tree, DarkNet-19, GoogLeNet, VGG, ResNet-50, Early-Exit ResNet, and
Qwen2.5-0.5B. No pretrained weights or DL-framework training are required.

\subsection{Installation}

The artifact is distributed as a Python codebase with JSON inputs and shell
scripts. All commands should run from the artifact root directory.
Install the Python packages and optional plotting dependency as follows:

\begin{verbatim}
cd sloth
pip install -r requirements.txt
pip install matplotlib
\end{verbatim}

\subsubsection{Installation validation}

After installing dependencies, run the smoke test to validate that the
simulator, trace writer, compressor, and RCA pipeline are working:
\begin{verbatim}
bash scripts/simulation.sh normal
bash scripts/simulation.sh fail
bash scripts/analysis.sh
\end{verbatim}
\verb|trace/<workload>/<arch>/<case>/{raw,compre|
\verb|-ssed}|
and \verb|trace/result/...| written the outputs. The RCA report is a JSON list of predicted
fail-slow components.
The report should list a PE that matches the injected component in \verb|data/fail_example.json|;
the metrics files should contain raw trace sizes and smaller compressed
trace sizes.

\subsection{Experiment workflow}

The workflow has two steps. First, generate execution
traces for different workloads on different accelerator architectures. Second,
run the experiment scripts based on the generated traces.

\subsubsection{Step 1: Generate execution traces}

Run the following command to sweep all supported workload, architecture, and fail-slow
configurations:
\begin{verbatim}
bash scripts/generate_traces.sh
\end{verbatim}
A reduced AE trace set can be generated with:
\begin{verbatim}
WORKLOAD_LIST="darknet19 resnet50" \
FAILSLOW_SET_LIST="center uniform" \
ARCH_LIST="mesh4_4" MAX_FAILS_PER_SET=1 \
PARALLEL_JOBS=1 \
bash scripts/generate_traces.sh
\end{verbatim}
Traces are organized by workload, architecture, and fail-slow case.
The \verb|normal| case is the no-fail-slow baseline for a workload and
architecture. Other cases correspond to injected fail-slow configurations.
Each case contains raw communication/computation traces and compressed traces
used by the detector.

\subsubsection{Step 2: Run experiments}

After traces are generated, run bash or Python scripts to produce
the paper results. The repository contains scripts for root-cause detection,
overhead analysis, SL-Recorder design-space exploration, severity analysis,
scalability analysis, baseline comparison. The specific
commands and expected results for these experiments are described in the next
section.

\subsection{Evaluation and expected results}

All commands should be run after trace generation unless the experiment script
explicitly generates its own traces. The plotting and table scripts read the
generated traces or metrics and save their outputs in \verb|figures/|.

\subsubsection{Detection accuracy and FPR comparison (Table 2)}

This experiment compares SLOTH with the baseline detectors using root-cause
detection accuracy and false-positive rate (FPR). Run the commands below:
\begin{verbatim}
bash scripts/analyze_traces.sh
bash scripts/run_baseline.sh
python3 draw/acc-tab-2.py
\end{verbatim}
\verb|analyze_traces.sh| runs SLOTH RCA and writes RCA reports. \verb|run_baseline.sh| runs Z-Thres, Mscope, IASO,
Perseus, and ADR and writes reports plus \verb|baseline_results.csv|.
\verb|acc-tab-2.py| formats Table~2 from the SLOTH and baseline metrics.
Reviewers should compare the workload-level accuracy and FPR entries produced
from the SLOTH \verb|overall.json| files with the corresponding baseline CSV
entries. The expected result is that SLOTH preserves the paper's accuracy/FPR
advantage over the baselines across the evaluated workloads.

\subsubsection{Overhead analysis (Figures 10--12)}

The overhead analysis measures execution-time overhead and trace-memory
overhead. Run:
\begin{verbatim}
python3 draw/overhead-fig-10.py
python3 draw/overhead-fig-11.py
python3 draw/overhead-fig-12.py
\end{verbatim}
\verb|overhead-fig-10.py| processes raw traces to compare baseline,
communication-probe, computation-probe, and full-probe execution time.
\verb|overhead-fig-11.py| and \verb|overhead-fig-12.py| compare raw and
compressed communication/computation trace sizes. Expected results are low
normalized execution-time overhead and much smaller compressed traces.
Figure~10 normalizes probe execution time to the no-probe baseline for each
workload. Figures~11 and 12 separate communication and computation traces so
reviewers can inspect whether the compression benefit holds for both parts of
the execution trace.

\subsubsection{DSE experiment (Figures 13--14)}

The DSE experiment explores SL-Recorder hash count, bucket count, stage size,
and threshold, and evaluates the accuracy/trace-memory trade-off. Run:
\begin{verbatim}
bash scripts/sketch_dse.sh
python3 draw/dse-fig-13.py
python3 draw/dse-fig-14.py
\end{verbatim}
\verb|sketch_dse.sh| sweeps recorder parameters, generates traces, runs RCA,
writes metrics plus \verb|results.txt| summaries.
\verb|dse-fig-13.py| plots pairwise parameter sensitivity. \verb|dse-fig-14.py|
plots the accuracy \& trace-memory trade-off for low-cost parameter settings.
Each DSE point corresponds to one recorder configuration and includes both
detection metrics and recorder memory metrics. Figure~13 shows how changing two
parameters at a time affects compression, while Figure~14 highlights the
settings that retain detection quality with lower trace-memory cost.

\subsubsection{Sensitivity experiment (Figure 15)}

The sensitivity experiment evaluates accuracy and FPR across slowdown ratios,
including 2x, 5x, and 10x. Run:
\begin{verbatim}
bash scripts/sensitivity.sh
python3 draw/sensitivity-fig-15.py
\end{verbatim}
\verb|sensitivity.sh| generates traces and prints RCA metrics for each slowdown ratio,
including \verb|overall.json| files and \verb|summary.tsv|.

\noindent \verb|sensitivity-fig-15.py| plots how accuracy and FPR vary over different
slowdown ratios. The expected result is that SLOTH remains effective when the
same fail-slow pattern is injected with different severity levels, while the
baseline curves remain below SLOTH.

\subsubsection{Scalability analysis (Figures 16--17)}

The scalability analysis evaluates SLOTH as accelerator scale increases across
4$\times$4, 6$\times$6, 8$\times$8, and 16$\times$16 mesh accelerators. Run:
\begin{verbatim}
python3 draw/scalability-fig-16.py
bash scripts/scalability_detect.sh
python3 draw/scalability-fig-17.py
\end{verbatim}
\verb|scalability-fig-16.py| plots execution-time overhead from the generated
scale traces. \verb|scalability_detect.sh| computes per-scale RCA metrics for
Figure~17. \verb|scalability-fig-17.py| combines trace-memory metrics and
accuracy results. Expected results are stable overhead and accuracy with reduced
trace size as accelerator scale increases.
Figure~16 uses the generated scale traces to estimate probe overhead as the
number of PEs and links grows. Figure~17 adds RCA results to show that the
compressed traces remain small while detection accuracy stays stable across
larger mesh accelerators.

\subsection{Notes}

Some experiments, especially full trace generation and DSE, are
time-consuming. The artifact provides cached traces, metrics, and plotting
inputs so reviewers can perform a fast evaluation without rerunning every experiment from scratch.
Some reproduced numbers may differ slightly from the paper because the
simulator uses a Gaussian performance variance model. These small differences
do not affect the trends or the main conclusions of the paper.

\bibliographystyle{IEEEtranS}
\bibliography{references}

@article{lie2023cerebras,
  title={Cerebras architecture deep dive: First look inside the hardware/software co-design for deep learning},
  author={Lie, Sean},
  journal={IEEE Micro},
  volume={43},
  number={3},
  pages={18--30},
  year={2023},
  publisher={IEEE}
}

@inproceedings{10.5555/3768039.3768105,
author = {Dong, Gen and Hua, Yu and Zhang, Yongle and Chen, Zhangyu and Chen, Menglei},
title = {Understanding and detecting fail-slow hardware failure bugs in cloud systems},
year = {2025},
isbn = {978-1-939133-48-9},
publisher = {USENIX Association},
address = {USA},
booktitle = {Proceedings of the 2025 USENIX Conference on Usenix Annual Technical Conference},
articleno = {66},
numpages = {16},
location = {Boston, MA, USA},
series = {USENIX ATC '25}
}

@article{gunawi2018fail,
  title={Fail-slow at scale: Evidence of hardware performance faults in large production systems},
  author={Gunawi, Haryadi S and Suminto, Riza O and Sears, Russell and Golliher, Casey and Sundararaman, Swaminathan and Lin, Xing and Emami, Tim and Sheng, Weiguang and Bidokhti, Nematollah and McCaffrey, Caitie and others},
  journal={ACM Transactions on Storage (TOS)},
  volume={14},
  number={3},
  pages={1--26},
  year={2018},
  publisher={ACM New York, NY, USA}
}

@inproceedings{panda2019iaso,
  title={$\{$IASO$\}$: A $\{$Fail-Slow$\}$ Detection and Mitigation Framework for Distributed Storage Services},
  author={Panda, Biswaranjan and Srinivasan, Deepthi and Ke, Huan and Gupta, Karan and Khot, Vinayak and Gunawi, Haryadi S},
  booktitle={2019 USENIX Annual Technical Conference (USENIX ATC 19)},
  pages={47--62},
  year={2019},
  publisher={USENIX Association},
  address={Renton, WA}
}

@inproceedings{lu2022nvme,
  title={$\{$NVMe$\}$$\{$SSD$\}$ failures in the field: the $\{$Fail-Stop$\}$ and the $\{$Fail-Slow$\}$},
  author={Lu, Ruiming and Xu, Erci and Zhang, Yiming and Zhu, Zhaosheng and Wang, Mengtian and Zhu, Zongpeng and Xue, Guangtao and Li, Minglu and Wu, Jiesheng},
  booktitle={2022 USENIX Annual Technical Conference (USENIX ATC 22)},
  pages={1005--1020},
  year={2022},
  publisher={USENIX Association},
  address={Carlsbad, CA, USA}
}

@INPROCEEDINGS{7900006,
  author={Teerapittayanon, Surat and McDanel, Bradley and Kung, H.T.},
  booktitle={2016 23rd International Conference on Pattern Recognition (ICPR)}, 
  title={BranchyNet: Fast inference via early exiting from deep neural networks}, 
  year={2016},
  volume={},
  number={},
  pages={2464-2469},
  doi={10.1109/ICPR.2016.7900006}}

@misc{hui2024qwen25codertechnicalreport,
      title={Qwen2.5-Coder Technical Report}, 
      author={Binyuan Hui and Jian Yang and Zeyu Cui and Jiaxi Yang and Dayiheng Liu and Lei Zhang and Tianyu Liu and Jiajun Zhang and Bowen Yu and Keming Lu and Kai Dang and Yang Fan and Yichang Zhang and An Yang and Rui Men and Fei Huang and Bo Zheng and Yibo Miao and Shanghaoran Quan and Yunlong Feng and Xingzhang Ren and Xuancheng Ren and Jingren Zhou and Junyang Lin},
      year={2024},
      eprint={2409.12186},
      archivePrefix={arXiv},
      primaryClass={cs.CL},
      url={https://arxiv.org/abs/2409.12186}, 
}

@inproceedings{10.5555/3585938.3585942,
    author = {Lu, Ruiming and Xu, Erci and Zhang, Yiming and Zhu, Fengyi and Zhu, Zhaosheng and Wang, Mengtian and Zhu, Zongpeng and Xue, Guangtao and Shu, Jiwu and Li, Minglu and Wu, Jiesheng},
    title = {PERSEUS: a fail-slow detection framework for cloud storage systems},
    year = {2023},
    isbn = {978-1-939133-32-8},
    publisher = {USENIX Association},
    address = {USA},
    booktitle = {Proceedings of the 21st USENIX Conference on File and Storage Technologies},
    articleno = {4},
    numpages = {15},
    location = {Santa Clara, CA, USA},
    series = {FAST'23}
}

@inproceedings{he2023understanding,
  title={Understanding and mitigating hardware failures in deep learning training systems},
  author={He, Yi and Hutton, Mike and Chan, Steven and De Gruijl, Robert and Govindaraju, Rama and Patil, Nishant and Li, Yanjing},
  booktitle={Proceedings of the 50th Annual International Symposium on Computer Architecture},
  pages={1--16},
  year={2023},
  publisher={ACM},
  address={FL, Orlando, USA}
}

@inproceedings{sastry2009mswat,
  title={mSWAT: Low-cost hardware fault detection and diagnosis for multicore systems},
  author={Sastry Hari, Siva Kumar and Li, Man-Lap and Ramachandran, Pradeep and Choi, Byn and Adve, Sarita V},
  booktitle={Proceedings of the 42nd Annual IEEE/ACM International Symposium on Microarchitecture},
  pages={122--132},
  year={2009},
  publisher={IEEE},
  address={New York, NY, USA}
}

@inproceedings{10.1145/3581784.3627042,
    author = {Ltaief, Hatem and Hong, Yuxi and Wilson, Leighton and Jacquelin, Mathias and Ravasi, Matteo and Keyes, David Elliot},
    title = {Scaling the “Memory Wall” for Multi-Dimensional Seismic Processing with Algebraic Compression on Cerebras CS-2 Systems},
    year = {2023},
    isbn = {9798400701092},
    publisher = {Association for Computing Machinery},
    address = {New York, NY, USA},
    url = {https://doi.org/10.1145/3581784.3627042},
    doi = {10.1145/3581784.3627042},
    booktitle = {Proceedings of the International Conference for High Performance Computing, Networking, Storage and Analysis},
    articleno = {6},
    numpages = {12},
    location = {Denver, CO, USA},
    series = {SC '23}
}

@misc{kundu2025comparisoncerebraswaferscaleintegration,
      title={A Comparison of the Cerebras Wafer-Scale Integration Technology with Nvidia GPU-based Systems for Artificial Intelligence}, 
      author={Yudhishthira Kundu and Manroop Kaur and Tripty Wig and Kriti Kumar and Pushpanjali Kumari and Vivek Puri and Manish Arora},
      year={2025},
      eprint={2503.11698},
      archivePrefix={arXiv},
      primaryClass={cs.AR},
      url={https://arxiv.org/abs/2503.11698}, 
}

@inproceedings{chang2011design,
  title={On the design and analysis of fault tolerant NoC architecture using spare routers},
  author={Chang, Yung-Chang and Chiu, Ching-Te and Lin, Shih-Yin and Liu, Chung-Kai},
  booktitle={16th Asia and South Pacific Design Automation Conference (ASP-DAC 2011)},
  pages={431--436},
  year={2011},
  organization={IEEE},
  publisher={IEEE},
  address={Yokohama, Japan}
}

@article{tsoutsouras2017softrm,
  title={SoftRM: Self-organized fault-tolerant resource management for failure detection and recovery in NoC based many-cores},
  author={Tsoutsouras, Vasileios and Masouros, Dimosthenis and Xydis, Sotirios and Soudris, Dimitrios},
  journal={ACM Transactions on Embedded Computing Systems (TECS)},
  volume={16},
  number={5s},
  pages={1--19},
  year={2017},
  publisher={ACM New York, NY, USA}
}

@inproceedings{taheri2022deft,
  title={DeFT: A deadlock-free and fault-tolerant routing algorithm for 2.5 D chiplet networks},
  author={Taheri, Ebadollah and Pasricha, Sudeep and Nikdast, Mahdi},
  booktitle={2022 Design, Automation \& Test in Europe Conference \& Exhibition (DATE)},
  pages={1047--1052},
  year={2022},
  organization={IEEE},
  publisher={IEEE},
  address={Antwerp, Belgium}
}

@ARTICLE{9076333,
  author={Kwon, Hyoukjun and Chatarasi, Prasanth and Sarkar, Vivek and Krishna, Tushar and Pellauer, Michael and Parashar, Angshuman},
  journal={IEEE Micro}, 
  title={MAESTRO: A Data-Centric Approach to Understand Reuse, Performance, and Hardware Cost of DNN Mappings}, 
  year={2020},
  volume={40},
  number={3},
  pages={20-29},
  doi={10.1109/MM.2020.2985963}
}

@INPROCEEDINGS{7056026,
  author={Wang, Xiaodong and Martínez, José F.},
  booktitle={2015 IEEE 21st International Symposium on High Performance Computer Architecture (HPCA)}, 
  title={XChange: A market-based approach to scalable dynamic multi-resource allocation in multicore architectures}, 
  year={2015},
  volume={},
  number={},
  pages={113-125},
  doi={10.1109/HPCA.2015.7056026},
  publisher={IEEE},
  address={Burlingame, CA, USA}
}

@INPROCEEDINGS{10485099,
  author={Wang, Pengyu and Yang, Weiling and Fang, Jianbin and Dong, Dezun and Huang, Chun and Zhang, Peng and Tang, Tao and Wang, Zheng},
  booktitle={SC23: International Conference for High Performance Computing, Networking, Storage and Analysis}, 
  title={Optimizing Direct Convolutions on ARM Multi-Cores}, 
  year={2023},
  volume={},
  number={},
  pages={1-14},
  doi={},
  publisher={ACM},
  address={CO, Denver, USA}
}

@article{chen2016eyeriss,
  title={Eyeriss: A spatial architecture for energy-efficient dataflow for convolutional neural networks},
  author={Chen, Yu-Hsin and Emer, Joel and Sze, Vivienne},
  journal={ACM SIGARCH computer architecture news},
  volume={44},
  number={3},
  pages={367--379},
  year={2016},
  publisher={ACM New York, NY, USA}
}

@inproceedings{hegde2021mind,
  title={Mind mappings: enabling efficient algorithm-accelerator mapping space search},
  author={Hegde, Kartik and Tsai, Po-An and Huang, Sitao and Chandra, Vikas and Parashar, Angshuman and Fletcher, Christopher W},
  booktitle={Proceedings of the 26th ACM International Conference on Architectural Support for Programming Languages and Operating Systems},
  pages={943--958},
  year={2021},
  publisher={ACM},
  address={Virtual, USA}
}

@inproceedings{tang2018vsensor,
  title={vSensor: leveraging fixed-workload snippets of programs for performance variance detection},
  author={Tang, Xiongchao and Zhai, Jidong and Qian, Xuehai and He, Bingsheng and Xue, Wei and Chen, Wenguang},
  booktitle={Proceedings of the 23rd ACM SIGPLAN symposium on principles and practice of parallel programming},
  pages={124--136},
  year={2018},
  publisher={ACM},
  address={Vienna Austria}
}

@inproceedings{jin2020scalana,
  title={ScalAna: Automating scaling loss detection with graph analysis},
  author={Jin, Yuyang and Wang, Haojie and Yu, Teng and Tang, Xiongchao and Hoefler, Torsten and Liu, Xu and Zhai, Jidong},
  booktitle={SC20: International Conference for High Performance Computing, Networking, Storage and Analysis},
  pages={1--14},
  year={2020},
  organization={IEEE},
  publisher={IEEE},
  address={Atlanta, GA, USA}
}

@inproceedings{zheng2022vapro,
  title={Vapro: Performance variance detection and diagnosis for production-run parallel applications},
  author={Zheng, Liyan and Zhai, Jidong and Tang, Xiongchao and Wang, Haojie and Yu, Teng and Jin, Yuyang and Song, Shuaiwen Leon and Chen, Wenguang},
  booktitle={Proceedings of the 27th ACM SIGPLAN Symposium on Principles and Practice of Parallel Programming},
  pages={150--162},
  year={2022},
  publisher={ACM},
  address={Seoul, Republic of Korea}
}

@inproceedings{Srinivasan2003RAMPA,
  title={RAMP : A Model for Reliability Aware MicroProcessor Design},
  author={Jayanth Srinivasan and Sarita V. Adve and Pradip Bose and Chao-Kun Hu},
  year={2003},
  url={https://api.semanticscholar.org/CorpusID:12656137}
}

@article{wu2014wafer,
  title={Wafer map failure pattern recognition and similarity ranking for large-scale data sets},
  author={Wu, Ming-Ju and Jang, Jyh-Shing R and Chen, Jui-Long},
  journal={IEEE Transactions on Semiconductor Manufacturing},
  volume={28},
  number={1},
  pages={1--12},
  year={2014},
  publisher={IEEE}
}

@inproceedings{hanson2023si,
  title={Si-kintsugi: Towards recovering golden-like performance of defective many-core spatial architectures for ai},
  author={Hanson, Edward and Li, Shiyu and Zhou, Guanglei and Cheng, Feng and Wang, Yitu and Bose, Rohan and Li, Hai and Chen, Yiran},
  booktitle={Proceedings of the 56th Annual IEEE/ACM International Symposium on Microarchitecture},
  pages={972--985},
  year={2023},
  publisher={ACM},
  address={ON, Toronto, Canada}
}

@software{simpy,
  author    = {K. G. Müller and T.Vignaux and O. Lünsdorf and S. Scherfke},
  title     = {SimPy: Discrete Event Simulation for Python},
  version   = {4.1.1},
  year      = {2002},
  url       = {https://simpy.readthedocs.io/},
  note      = {Accessed: 2023-11-13},
  organization={Team SimPy}
}

@inproceedings{10.1109/ISCA59077.2024.00046,
    author = {Zhang, Nathan and Lacouture, Rubens and Sohn, Gina and Mure, Paul and Zhang, Qizheng and Kjolstad, Fredrik and Olukotun, Kunle},
    title = {The Dataflow Abstract Machine Simulator Framework},
    year = {2024},
    isbn = {9798350326581},
    publisher = {IEEE Press},
    url = {https://doi.org/10.1109/ISCA59077.2024.00046},
    doi = {10.1109/ISCA59077.2024.00046},
    booktitle = {Proceedings of the 51st Annual International Symposium on Computer Architecture},
    pages = {532–547},
    numpages = {16},
    address = {Buenos Aires, Argentina},
    series = {ISCA '24}
}

@article{pillai2017application,
  title={Application crash consistency and performance with CCFS},
  author={Pillai, Thanumalayan Sankaranarayana and Alagappan, Ramnatthan and Lu, Lanyue and Chidambaram, Vijay and Arpaci-Dusseau, Andrea C and Arpaci-Dusseau, Remzi H},
  journal={ACM Transactions on Storage (TOS)},
  volume={13},
  number={3},
  pages={1--29},
  year={2017},
  publisher={ACM New York, NY, USA}
}

@inproceedings{alagappan2016correlated,
  title={Correlated crash vulnerabilities},
  author={Alagappan, Ramnatthan and Ganesan, Aishwarya and Patel, Yuvraj and Pillai, Thanumalayan Sankaranarayana and Arpaci-Dusseau, Andrea C and Arpaci-Dusseau, Remzi H},
  booktitle={12th USENIX Symposium on Operating Systems Design and Implementation (OSDI 16)},
  pages={151--167},
  year={2016},
  publisher={USENIX Association},
  address={Savannah, GA, USA}
}

@inproceedings{haghbayan2020thermal,
  title={Thermal-cycling-aware dynamic reliability management in many-core system-on-chip},
  author={Haghbayan, Mohammad-Hashem and Miele, Antonio and Zou, Zhuo and Tenhunen, Hannu and Plosila, Juha},
  booktitle={2020 Design, Automation \& Test in Europe Conference \& Exhibition (DATE)},
  pages={1229--1234},
  year={2020},
  organization={IEEE},
  publisher={IEEE},
  address={Grenoble, France}
}

@article{wang2019optimized,
  title={Optimized mapping algorithm to extend lifetime of both NoC and cores in many-core system},
  author={Wang, Lihuan and Jiang, Shuyan and Chen, Shuyu and Wang, Junshi and Huang, Letian},
  journal={Integration},
  volume={67},
  pages={82--94},
  year={2019},
  publisher={Elsevier}
}

@inproceedings{agarwal2007circuit,
  title={Circuit failure prediction and its application to transistor aging},
  author={Agarwal, Mridul and Paul, Bipul C and Zhang, Ming and Mitra, Subhasish},
  booktitle={25th IEEE VLSI Test Symposium (VTS'07)},
  pages={277--286},
  year={2007},
  organization={IEEE},
  publisher={IEEE},
  address = {Berkeley, CA, USA}
}

@inproceedings{bhardwaj2012towards,
  title={Towards graceful aging degradation in NoCs through an adaptive routing algorithm},
  author={Bhardwaj, Kshitij and Chakraborty, Koushik and Roy, Sanghamitra},
  booktitle={Proceedings of the 49th Annual Design Automation Conference},
  pages={382--391},
  year={2012},
  publisher={ACM},
  address={San Francisco, California USA}
}

@article{lin2017thermal,
  title={Thermal-and performance-aware address mapping for the multi-channel three-dimensional DRAM systems},
  author={Lin, Shu-Yen and Lin, Jin-Yi},
  journal={IEEE Access},
  volume={5},
  pages={5566--5577},
  year={2017},
  publisher={IEEE}
}

@article{pandey20243d,
  title={3D-TemPo: Optimizing 3D DRAM performance under temperature and power constraints},
  author={Pandey, Shailja and Sethi, Sayam and Panda, Preeti Ranjan},
  journal={IEEE Transactions on Computer-Aided Design of Integrated Circuits and Systems},
  volume={43},
  number={8},
  pages={2263 - 2276},
  year={2024},
  publisher={IEEE}
}

@inproceedings{shukla2019overview,
  title={An overview of thermal challenges and opportunities for monolithic 3D ICs},
  author={Shukla, Prachi and Coskun, Ayse K and Pavlidis, Vasilis F and Salman, Emre},
  booktitle={Proceedings of the 2019 Great Lakes Symposium on VLSI},
  pages={439--444},
  year={2019},
  publisher={ACM},
  address={VA, Tysons Corner, USA}
}

@inproceedings{banakar2002scratchpad,
  title={Scratchpad memory: design alternative for cache on-chip memory in embedded systems},
  author={Banakar, Rajeshwari and Steinke, Stefan and Lee, Bo-Sik and Balakrishnan, Mahesh and Marwedel, Peter},
  booktitle={Proceedings of the tenth international symposium on Hardware/software codesign},
  pages={73--78},
  year={2002},
  publisher={ACM},
  address={Estes Park, Colorado, USA}
}

@article{nikhil2002executing,
  title={Executing a program on the MIT tagged-token dataflow architecture},
  author={Nikhil, Rishiyur S and others},
  journal={IEEE Transactions on computers},
  volume={39},
  number={3},
  pages={300--318},
  year={2002},
  publisher={IEEE}
}

@inproceedings{dennis1974preliminary,
  title={A preliminary architecture for a basic data-flow processor},
  author={Dennis, Jack B and Misunas, David P},
  booktitle={Proceedings of the 2nd annual symposium on Computer architecture},
  pages={126--132},
  year={1974},
  publisher={ACM},
  address={Barcelona, Spain}
}

@inproceedings{10.1145/1176887.1176933,
    author = {Egger, Bernhard and Lee, Jaejin and Shin, Heonshik},
    title = {Scratchpad memory management for portable systems with a memory management unit},
    year = {2006},
    isbn = {1595935428},
    publisher = {Association for Computing Machinery},
    address = {New York, NY, USA},
    url = {https://doi.org/10.1145/1176887.1176933},
    doi = {10.1145/1176887.1176933},
    booktitle = {Proceedings of the 6th ACM \& IEEE International Conference on Embedded Software},
    pages = {321–330},
    numpages = {10},
    location = {Seoul, Korea},
    series = {EMSOFT '06}
}

@inproceedings{gratz2006implementation,
  title={Implementation and evaluation of on-chip network architectures},
  author={Gratz, Paul and Kim, Changkyu and McDonald, Robert and Keckler, Stephen W and Burger, Doug},
  booktitle={2006 International Conference on Computer Design},
  pages={477--484},
  year={2006},
  organization={IEEE},
  publisher={IEEE},
  address={San Jose, CA, USA}
}

@inproceedings {276934,
  author = {Hun Namkung and Zaoxing Liu and Daehyeok Kim and Vyas Sekar and Peter Steenkiste},
  title = {{SketchLib}: Enabling Efficient Sketch-based Monitoring on Programmable Switches},
  booktitle = {19th USENIX Symposium on Networked Systems Design and Implementation (NSDI 22)},
  year = {2022},
  isbn = {978-1-939133-27-4},
  address = {Renton, WA},
  pages = {743--759},
  url = {https://www.usenix.org/conference/nsdi22/presentation/namkung},
  publisher = {USENIX Association},
  month = apr
}

@inproceedings{taylor2003scalar,
  title={Scalar operand networks: On-chip interconnect for ILP in partitioned architectures},
  author={Taylor, M Bedford and Lee, Walter and Amarasinghe, Saman and Agarwal, Anant},
  booktitle={The Ninth International Symposium on High-Performance Computer Architecture, 2003. HPCA-9 2003. Proceedings.},
  pages={341--353},
  year={2003},
  organization={IEEE},
  publisher={IEEE},
  address={Anaheim, CA, USA}
}

@inproceedings{zheng2021adapt,
  title={Adapt-noc: A flexible network-on-chip design for heterogeneous manycore architectures},
  author={Zheng, Hao and Wang, Ke and Louri, Ahmed},
  booktitle={2021 IEEE international symposium on high-performance computer architecture (HPCA)},
  pages={723--735},
  year={2021},
  organization={IEEE},
  publisher={IEEE},
  address={Seoul, Korea (South)}
}

@inproceedings{feng2024ring,
  title={Ring Road: A Scalable Polar-Coordinate-based 2D Network-on-Chip Architecture},
  author={Feng, Yinxiao and Li, Wei and Ma, Kaisheng},
  booktitle={2024 57th IEEE/ACM International Symposium on Microarchitecture (MICRO)},
  pages={871--884},
  year={2024},
  organization={IEEE},
  publisher={IEEE},
  address={Austin, TX, USA}
}

@inproceedings{10.5555/3767955.3767975,
author = {Lu, Ruiming and Lu, Yunchi and Jiang, Yuxuan and Xue, Guangtao and Huang, Peng},
title = {One-size-fits-none: understanding and enhancing slow-fault tolerance in modern distributed systems},
year = {2025},
isbn = {978-1-939133-46-5},
publisher = {USENIX Association},
address = {USA},
booktitle = {Proceedings of the 22nd USENIX Symposium on Networked Systems Design and Implementation},
articleno = {20},
numpages = {20},
location = {Philadelphia, PA, USA},
series = {NSDI '25}
}

@inproceedings{10.1145/3132747.3132749,
author = {Kaldor, Jonathan and Mace, Jonathan and Bejda, Micha\l{} and Gao, Edison and Kuropatwa, Wiktor and O'Neill, Joe and Ong, Kian Win and Schaller, Bill and Shan, Pingjia and Viscomi, Brendan and Venkataraman, Vinod and Veeraraghavan, Kaushik and Song, Yee Jiun},
title = {Canopy: An End-to-End Performance Tracing And Analysis System},
year = {2017},
isbn = {9781450350853},
publisher = {Association for Computing Machinery},
address = {New York, NY, USA},
url = {https://doi.org/10.1145/3132747.3132749},
doi = {10.1145/3132747.3132749},
booktitle = {Proceedings of the 26th Symposium on Operating Systems Principles},
pages = {34–50},
numpages = {17},
location = {Shanghai, China},
series = {SOSP '17}
}

@inproceedings{10.1145/3357223.3362736,
author = {Las-Casas, Pedro and Papakerashvili, Giorgi and Anand, Vaastav and Mace, Jonathan},
title = {Sifter: Scalable Sampling for Distributed Traces, without Feature Engineering},
year = {2019},
isbn = {9781450369732},
publisher = {Association for Computing Machinery},
address = {New York, NY, USA},
url = {https://doi.org/10.1145/3357223.3362736},
doi = {10.1145/3357223.3362736},
booktitle = {Proceedings of the ACM Symposium on Cloud Computing},
pages = {312–324},
numpages = {13},
location = {Santa Cruz, CA, USA},
series = {SoCC '19}
}

@inproceedings{10.1145/3267809.3267841,
author = {Las-Casas, Pedro and Mace, Jonathan and Guedes, Dorgival and Fonseca, Rodrigo},
title = {Weighted Sampling of Execution Traces: Capturing More Needles and Less Hay},
year = {2018},
isbn = {9781450360111},
publisher = {Association for Computing Machinery},
address = {New York, NY, USA},
url = {https://doi.org/10.1145/3267809.3267841},
doi = {10.1145/3267809.3267841},
booktitle = {Proceedings of the ACM Symposium on Cloud Computing},
pages = {326–332},
numpages = {7},
location = {Carlsbad, CA, USA},
series = {SoCC '18}
}

@inproceedings{10.1109/ASE.2019.00085,
author = {Liu, Jinyang and Zhu, Jieming and He, Shilin and He, Pinjia and Zheng, Zibin and Lyu, Michael R.},
title = {Logzip: extracting hidden structures via iterative clustering for log compression},
year = {2020},
isbn = {9781728125084},
publisher = {IEEE Press},
url = {https://doi.org/10.1109/ASE.2019.00085},
doi = {10.1109/ASE.2019.00085},
booktitle = {Proceedings of the 34th IEEE/ACM International Conference on Automated Software Engineering},
pages = {863–873},
numpages = {11},
location = {San Diego, California},
series = {ASE '19}
}

@article{10.1109/TKDE.2022.3223686,
author = {Miao, Ruijie and Zhong, Zheng and Guo, Jiarui and Li, Zikun and Yang, Tong and Cui, Bin},
title = {BurstSketch: Finding Bursts in Data Streams},
year = {2023},
issue_date = {Nov. 2023},
publisher = {IEEE Educational Activities Department},
address = {USA},
volume = {35},
number = {11},
issn = {1041-4347},
url = {https://doi.org/10.1109/TKDE.2022.3223686},
doi = {10.1109/TKDE.2022.3223686},
journal = {IEEE Trans. on Knowl. and Data Eng.},
month = nov,
pages = {11126–11140},
numpages = {15}
}

@inproceedings {273737,
author = {Kirk Rodrigues and Yu Luo and Ding Yuan},
title = {{CLP}: Efficient and Scalable Search on Compressed Text Logs},
booktitle = {15th {USENIX} Symposium on Operating Systems Design and Implementation ({OSDI} 21)},
year = {2021},
isbn = {978-1-939133-22-9},
pages = {183--198},
url = {https://www.usenix.org/conference/osdi21/presentation/rodrigues},
publisher = {{USENIX} Association},
month = jul
}

@techreport{36356,title	= {Dapper, a Large-Scale Distributed Systems Tracing Infrastructure},author	= {Benjamin H. Sigelman and Luiz André Barroso and Mike Burrows and Pat Stephenson and Manoj Plakal and Donald Beaver and Saul Jaspan and Chandan Shanbhag},year	= {2010},URL	= {http://research.google.com/archive/papers/dapper-2010-1.pdf},institution	= {Google, Inc.}}

@inproceedings{feng2023scalable,
  title={A scalable methodology for designing efficient interconnection network of chiplets},
  author={Feng, Yinxiao and Xiang, Dong and Ma, Kaisheng},
  booktitle={2023 IEEE International Symposium on High-Performance Computer Architecture (HPCA)},
  pages={1059--1071},
  year={2023},
  organization={IEEE},
  publisher={IEEE},
  address={Montreal, QC, Canada}
}

@inproceedings{marathe2017empirical,
  title={An empirical survey of performance and energy efficiency variation on intel processors},
  author={Marathe, Aniruddha and Zhang, Yijia and Blanks, Grayson and Kumbhare, Nirmal and Abdulla, Ghaleb and Rountree, Barry},
  booktitle={Proceedings of the 5th International Workshop on Energy Efficient Supercomputing},
  pages={1--8},
  year={2017},
  publisher={ACM},
  address={CO, Denver, USA}
}

@inproceedings{inadomi2015analyzing,
  title={Analyzing and mitigating the impact of manufacturing variability in power-constrained supercomputing},
  author={Inadomi, Yuichi and Patki, Tapasya and Inoue, Koji and Aoyagi, Mutsumi and Rountree, Barry and Schulz, Martin and Lowenthal, David and Wada, Yasutaka and Fukazawa, Keiichiro and Ueda, Masatsugu and others},
  booktitle={Proceedings of the international conference for high performance computing, networking, storage and analysis},
  pages={1--12},
  year={2015},
  publisher={ACM},
  address={Texas, Austin}
}

@inproceedings{romanescu2006quantifying,
  title={Quantifying the impact of process variability on microprocessor behavior},
  author={Romanescu, Bogdan F and Ozev, Sule and Sorin, Daniel J},
  booktitle={Workshop on Architectural Reliability},
  numpages={10},
  year={2006},
  organization={Citeseer},
  address={Orlando, Florida}
}

@inproceedings{hernandez2010methodology,
  title={A methodology for the characterization of process variation in NoC links},
  author={Hern{\'a}ndez, Carles and Silla, Federico and Duato, Jos{\'e}},
  booktitle={2010 Design, Automation \& Test in Europe Conference \& Exhibition (DATE 2010)},
  pages={685--690},
  year={2010},
  organization={IEEE},
  publisher={IEEE},
  address={Dresden, Germany}
}

@inproceedings{cai2024gemini,
  title={Gemini: Mapping and architecture co-exploration for large-scale dnn chiplet accelerators},
  author={Cai, Jingwei and Wu, Zuotong and Peng, Sen and Wei, Yuchen and Tan, Zhanhong and Shi, Guiming and Gao, Mingyu and Ma, Kaisheng},
  booktitle={2024 IEEE International Symposium on High-Performance Computer Architecture (HPCA)},
  pages={156--171},
  year={2024},
  organization={IEEE},
  publisher={IEEE},
  address={Edinburgh, United Kingdom}
}

@inproceedings{szegedy2015going,
  title={Going deeper with convolutions},
  author={Szegedy, Christian and Liu, Wei and Jia, Yangqing and Sermanet, Pierre and Reed, Scott and Anguelov, Dragomir and Erhan, Dumitru and Vanhoucke, Vincent and Rabinovich, Andrew},
  booktitle={Proceedings of the IEEE conference on computer vision and pattern recognition},
  pages={1--9},
  year={2015},
  publisher={IEEE},
  address={Boston, MA, USA}
}

@INPROCEEDINGS{8100173,
  author={Redmon, Joseph and Farhadi, Ali},
  booktitle={2017 IEEE Conference on Computer Vision and Pattern Recognition (CVPR)}, 
  title={YOLO9000: Better, Faster, Stronger}, 
  year={2017},
  volume={},
  number={},
  pages={6517-6525},
  doi={10.1109/CVPR.2017.690},
  publisher={IEEE},
  address={Honolulu, HI, USA}
}

@misc{simonyan2015deepconvolutionalnetworkslargescale,
      title={Very Deep Convolutional Networks for Large-Scale Image Recognition}, 
      author={Karen Simonyan and Andrew Zisserman},
      year={2015},
      eprint={1409.1556},
      archivePrefix={arXiv},
      primaryClass={cs.CV},
      url={https://arxiv.org/abs/1409.1556}, 
}

@INPROCEEDINGS{7780459,
  author={He, Kaiming and Zhang, Xiangyu and Ren, Shaoqing and Sun, Jian},
  booktitle={2016 IEEE Conference on Computer Vision and Pattern Recognition (CVPR)}, 
  title={Deep Residual Learning for Image Recognition}, 
  year={2016},
  volume={},
  number={},
  pages={770-778},
  doi={10.1109/CVPR.2016.90},
  publisher={IEEE},
  address={Las Vegas, NV, USA}
}

@inproceedings{sage,
  author = {Gan, Yu and Liang, Mingyu and Dev, Sundar and Lo, David and Delimitrou, Christina},
  title = {Sage: practical and scalable ML-driven performance debugging in microservices},
  year = {2021},
  isbn = {9781450383172},
  publisher = {Association for Computing Machinery},
  address = {New York, NY, USA},
  url = {https://doi.org/10.1145/3445814.3446700},
  doi = {10.1145/3445814.3446700},
  booktitle = {Proceedings of the 26th ACM International Conference on Architectural Support for Programming Languages and Operating Systems},
  pages = {135–151},
  numpages = {17},
  location = {Virtual, USA},
  series = {ASPLOS '21}
}

@inproceedings{sleuth,
    author = {Gan, Yu and Liu, Guiyang and Zhang, Xin and Zhou, Qi and Wu, Jiesheng and Jiang, Jiangwei},
    title = {Sleuth: A Trace-Based Root Cause Analysis System for Large-Scale Microservices with Graph Neural Networks},
    year = {2024},
    isbn = {9798400703942},
    publisher = {Association for Computing Machinery},
    address = {New York, NY, USA},
    url = {https://doi.org/10.1145/3623278.3624758},
    doi = {10.1145/3623278.3624758},
    booktitle = {Proceedings of the 28th ACM International Conference on Architectural Support for Programming Languages and Operating Systems, Volume 4},
    pages = {324–337},
    numpages = {14},
    location = {Vancouver, BC, Canada},
    series = {ASPLOS '23}
}

@misc{wu2024falconpinpointingmitigatingstragglers,
      title={FALCON: Pinpointing and Mitigating Stragglers for Large-Scale Hybrid-Parallel Training}, 
      author={Tianyuan Wu and Wei Wang and Yinghao Yu and Siran Yang and Wenchao Wu and Qinkai Duan and Guodong Yang and Jiamang Wang and Lin Qu and Liping Zhang},
      year={2024},
      eprint={2410.12588},
      archivePrefix={arXiv},
      primaryClass={cs.DC},
      url={https://arxiv.org/abs/2410.12588}, 
}

@inproceedings{groot,
author = {Wang, Hanzhang and Wu, Zhengkai and Jiang, Huai and Huang, Yichao and Wang, Jiamu and Kopru, Selcuk and Xie, Tao},
title = {Groot: an event-graph-based approach for root cause analysis in industrial settings},
year = {2022},
isbn = {9781665403375},
publisher = {IEEE Press},
url = {https://doi.org/10.1109/ASE51524.2021.9678708},
doi = {10.1109/ASE51524.2021.9678708},
booktitle = {Proceedings of the 36th IEEE/ACM International Conference on Automated Software Engineering},
pages = {419–429},
numpages = {11},
location = {Melbourne, Australia},
series = {ASE '21},
address={Melbourne, Australia}
}

@article{lin2025understanding,
  title={Understanding Stragglers in Large Model Training Using What-if Analysis},
  author={Lin, Jinkun and Jiang, Ziheng and Song, Zuquan and Zhao, Sida and Yu, Menghan and Wang, Zhanghan and Wang, Chenyuan and Shi, Zuocheng and Shi, Xiang and Jia, Wei and others},
  journal={arXiv preprint arXiv:2505.05713},
  year={2025},
  numpages={19}
}

@inproceedings {280874,
    author = {Lianmin Zheng and Zhuohan Li and Hao Zhang and Yonghao Zhuang and Zhifeng Chen and Yanping Huang and Yida Wang and Yuanzhong Xu and Danyang Zhuo and Eric P. Xing and Joseph E. Gonzalez and Ion Stoica},
    title = {Alpa: Automating Inter- and {Intra-Operator} Parallelism for Distributed Deep Learning},
    booktitle = {16th USENIX Symposium on Operating Systems Design and Implementation (OSDI 22)},
    year = {2022},
    isbn = {978-1-939133-28-1},
    address = {Carlsbad, CA},
    pages = {559--578},
    url = {https://www.usenix.org/conference/osdi22/presentation/zheng-lianmin},
    publisher = {USENIX Association},
    month = jul
}

@inproceedings{10.1007/978-3-030-03596-9_1,
author = {Lin, Jinjin and Chen, Pengfei and Zheng, Zibin},
title = {Microscope: Pinpoint Performance Issues with Causal Graphs in Micro-service Environments},
year = {2018},
isbn = {978-3-030-03595-2},
publisher = {Springer-Verlag},
address = {Berlin, Heidelberg},
url = {https://doi.org/10.1007/978-3-030-03596-9_1},
doi = {10.1007/978-3-030-03596-9_1},
booktitle = {Service-Oriented Computing: 16th International Conference, ICSOC 2018, Hangzhou, China, November 12-15, 2018, Proceedings},
pages = {3–20},
numpages = {18},
location = {Hangzhou, China}
}

@inproceedings{10.1145/3669940.3707287,
author = {Huang, Haiyu and Chen, Cheng and Chen, Kunyi and Chen, Pengfei and Yu, Guangba and He, Zilong and Wang, Yilun and Zhang, Huxing and Zhou, Qi},
title = {Mint: Cost-Efficient Tracing with All Requests Collection via Commonality and Variability Analysis},
year = {2025},
isbn = {9798400706981},
publisher = {Association for Computing Machinery},
address = {New York, NY, USA},
url = {https://doi.org/10.1145/3669940.3707287},
doi = {10.1145/3669940.3707287},
booktitle = {Proceedings of the 30th ACM International Conference on Architectural Support for Programming Languages and Operating Systems, Volume 1},
pages = {683–697},
numpages = {15},
location = {Rotterdam, Netherlands},
series = {ASPLOS '25}
}

@inproceedings{10.1145/2815400.2815415,
author = {Mace, Jonathan and Roelke, Ryan and Fonseca, Rodrigo},
title = {Pivot tracing: dynamic causal monitoring for distributed systems},
year = {2015},
isbn = {9781450338349},
publisher = {Association for Computing Machinery},
address = {New York, NY, USA},
url = {https://doi.org/10.1145/2815400.2815415},
doi = {10.1145/2815400.2815415},
booktitle = {Proceedings of the 25th Symposium on Operating Systems Principles},
pages = {378–393},
numpages = {16},
location = {Monterey, California},
series = {SOSP '15}
}

@inproceedings {285115,
    author = {Lei Zhang and Zhiqiang Xie and Vaastav Anand and Ymir Vigfusson and Jonathan Mace},
    title = {The Benefit of Hindsight: Tracing {Edge-Cases} in Distributed Systems},
    booktitle = {20th USENIX Symposium on Networked Systems Design and Implementation (NSDI 23)},
    year = {2023},
    isbn = {978-1-939133-33-5},
    address = {Boston, MA},
    pages = {321--339},
    url = {https://www.usenix.org/conference/nsdi23/presentation/zhang-lei},
    publisher = {USENIX Association},
}

@INPROCEEDINGS{9590295,
  author={Huang, Zicheng and Chen, Pengfei and Yu, Guangba and Chen, Hongyang and Zheng, Zibin},
  booktitle={2021 IEEE International Conference on Web Services (ICWS)}, 
  title={Sieve: Attention-based Sampling of End-to-End Trace Data in Distributed Microservice Systems}, 
  year={2021},
  volume={},
  number={},
  pages={436-446},
  doi={10.1109/ICWS53863.2021.00063},
  publisher={IEEE},
  address={Chicago, IL, USA}
}

@INPROCEEDINGS{9064148,
  author={Wang, Yipeng and Yang, Tong and Wang, Ren and Tai, Charlie},
  booktitle={2019 IEEE 8th International Conference on Cloud Networking (CloudNet)}, 
  title={Dynamic Sketch: Efficient and Adjustable Heavy Hitter Detection for Software Packet Processing}, 
  year={2019},
  volume={},
  number={},
  pages={1-7},
  doi={10.1109/CloudNet47604.2019.9064148},
  publisher={IEEE},
  address={Coimbra, Portugal}
}

@inproceedings{10.1145/3589334.3645581,
author = {Li, Weihe and Patras, Paul},
title = {Stable-Sketch: A Versatile Sketch for Accurate, Fast, Web-Scale Data Stream Processing},
year = {2024},
isbn = {9798400701719},
publisher = {Association for Computing Machinery},
address = {New York, NY, USA},
url = {https://doi.org/10.1145/3589334.3645581},
doi = {10.1145/3589334.3645581},
booktitle = {Proceedings of the ACM Web Conference 2024},
pages = {4227–4238},
numpages = {12},
location = {Singapore, Singapore},
series = {WWW '24}
}

@article{tyrrell2024revisiting,
  title={Revisiting Reliability in Large-Scale Machine Learning Research Clusters},
  author={Tyrrell, Virginia and Phanishayee, Amar and Banerjee, Siddhartha and Jiang, Junchen},
  journal={arXiv preprint arXiv:2410.21680},
  year={2024}
}

@inproceedings{prabhakar2024sambanova,
  title={{SambaNova SN40L}: Scaling the {AI} Memory Wall with Dataflow and Composition of Experts},
  author={Prabhakar, Raghu and Sivaramakrishnan, Sumti and Panda, Ravi and Zhang, Cing-Yu and Zhang, Tian and Hong, Deming and Kaur, Jasjeet and Mao, Jiafei and Patil, Muralidhar and Rui, Jie and others},
  booktitle={Proceedings of the 57th IEEE/ACM International Symposium on Microarchitecture (MICRO)},
  year={2024}
}

@inproceedings{genc2024stellar,
  title={Stellar: An Automated Design Framework for Dense and Sparse Spatial Accelerators},
  author={Genc, Hasan and Kim, Seah and Ganesh, Aditya and Shao, Yakun Sophia},
  booktitle={Proceedings of the 57th IEEE/ACM International Symposium on Microarchitecture (MICRO)},
  year={2024}
}

\end{document}